\documentclass[fleqn,usenatbib]{mnras}

\usepackage{newtxtext,newtxmath}
 
\usepackage[T1]{fontenc}

\DeclareRobustCommand{\VAN}[3]{#2}
\let\VANthebibliography\thebibliography
\def\thebibliography{\DeclareRobustCommand{\VAN}[3]{##3}\VANthebibliography}

\usepackage{graphicx}	
\usepackage{amsmath}	
\usepackage{amssymb}	

\newcommand{\erasst}{eRASSt J234402.9-352640}

\title[Radio observations of J2344]{A radio flare associated with the nuclear transient eRASSt J234403-352640: an outflow launched by a potential tidal disruption event}

\author[A. J. Goodwin et. al.]{
A. J. Goodwin,$^{1}$\thanks{E-mail: ajgoodwin.astro@gmail.com} G. E. Anderson,$^{1}$ J. C. A. Miller-Jones,$^{1}$ A. Malyali,$^{2}$ I. Grotova,$^{2}$ D. Homan,$^{3}$ A. Kawka,$^{1}$
\newauthor
M. Krumpe,$^{3}$ Z. Liu,${^2}$ and A. Rau${^2}$
\\
$^{1}$International Centre for Radio Astronomy Research -- Curtin University, GPO Box U1987, Perth, WA 6845, Australia \\
$^{2}$Max-Planck-Institut f\"ur extraterrestrische Physik,  Giessenbachstrasse 1, 85748 Garching, Germany\\ 
$^{3}$ Leibniz-Institut f\"ur Astrophysik Potsdam, An der Sternwarte 16, 14482 Potsdam, Germany\\
}

\date{Accepted XXX. Received YYY; in original form ZZZ}

\pubyear{2023}

\begin{document}
\label{firstpage}
\pagerange{\pageref{firstpage}--\pageref{lastpage}}
\maketitle

\begin{abstract}
We present an extensive radio monitoring campaign of the nuclear transient \erasst{} with the Australia Telescope Compact Array, one of the most X-ray luminous TDE candidates discovered by the SRG/eROSITA all-sky survey. The observations reveal a radio flare lasting $>1000$\,d, coincident with the X-ray, UV, optical, and infra-red flare of this transient event. 
Through modelling of the 10 epochs of radio spectral observations obtained, we find that the radio emission is well-described by an expanding synchrotron emitting region, consisting of a single ejection of material launched coincident with the optical flare.
We conclude that the radio flare properties of \erasst{} are consistent with the population of radio-emitting outflows launched by non-relativistic tidal disruption events, and that the flare is likely due to an outflow launched by a tidal disruption event (but could also be a due to a new AGN accretion event) in a previously turned-off AGN. 
\end{abstract}




\section{Introduction}

The release of gravitational energy when mass is suddenly accreted onto a black hole powers some of the most explosive phenomena in the Universe. Most galactic nuclei host a supermassive black hole \citep[SMBH]{Soltan1982}, which can power highly energetic outflows from accretion events. Tidal disruption events are an extreme example of this process, occurring when a star passes too close to a supermassive black hole and is pulled apart by strong tidal forces \citep[e.g.][]{Hills1975,Rees1988}. Approximately 50\% of the stellar material is initially bound to the black hole and accretes at rates approaching or surpassing the Eddington limit over a short ($\sim$yrs) timescale, with energy being fed back into the surrounding galaxy via outflows and radiation. Extreme variability and flaring events can also be caused by a sudden enhancement in a more persistent accretion flow onto a SMBH powering an active galactic nucleus (AGN) \citep{Rees1984}, with AGN flaring behaviour thought to be due to sudden accretion episodes or magnetic field changes \citep[e.g.][]{deVries1992,Ciotti2007,Hovatta2008}. Therefore the observational signatures of AGN flares and TDEs are similar as they are likely driven by the same process of a sudden increase in accretion rate onto the SMBH, although differences in the evolution of the accretion rate and circumnuclear environment may affect the evolution of the flare in each case.



Following the stellar disruption during a TDE, the bound material is thought to be the source of observed optical and X-ray emission \citep[e.g.][]{vanVelzen2020}, whilst the unbound, ejected material is thought to produce synchrotron emission from shocks with the surrounding circumnuclear medium (CNM), detected at radio wavelengths \citep[e.g.][]{Alexander2020}. The first observed TDE candidates were discovered by the ROSAT All Sky Survey \citep{Truemper1982} as luminous (10$^{41-44}$\,erg\,s$^{-1}$) X-ray outbursts with extremely soft ($\Gamma>3$) X-ray spectra in galaxies that had no previous AGN activity  \citep[e.g.][]{Komossa&Bade1999,Donley2002}.
Optically-discovered TDEs radiate much of their energy in the optical/UV band \citep[e.g.][]{vanVelzen2021} showing thermal blackbody emission with temperatures of $\sim10^{4}$\,K \citep{Gezari2009,vanVelzen2020}, although, when discovered at wavelengths other than optical/UV can be optically dim 
\citep[e.g.][]{Malyali2023,Saxton2020,Mattila2018}. Many TDE optical light curves decay at early times broadly consistent with a $t^{-5/3}$ decay, appearing to trace the theoretical rate of mass fall-back for complete disruptions \citep{Phinney1989,Guillochon2013}. There is a growing population of TDEs with unusual late-time optical behaviour, including rebrightenings \citep{Yao2023,Hammerstein2022}.
Thermal X-rays are detected only in some events, with the X-ray rise potentially significantly delayed from the initial optical flare \citep[e.g.][]{Kajava2020}. The X-ray decay phases can be highly non-monotonic \citep[see discussion in][]{Malyali2023}, in contrast with the relatively smoothly-declining optical lightcurves.
It has been argued that the diversity of observational properties in TDEs is due to the varying mass, type, density, or structure of the star that is disrupted, varying black hole mass and spin, the orbit of the disrupted star, the host galaxy properties (including the presence of a pre-existing accretion disk), and the viewing angle
\citep{Lodato2009,Guillochon2015,Dai2018,Malyali2023b,Liu2023,Wevers2023}.

Radio emission in TDEs is associated with synchrotron emission from outflowing material shocking the CNM of the host galaxy \citep[see][for a review]{Alexander2020}. Currently, the outflow mechanism is unknown but has been proposed to be accretion-powered jets or winds \citep[e.g.][]{Alexander2016,Stein2021,Cendes2022,Somalwar2022,Somalwar2023,Ravi2022}, collision induced outflows \citep[e.g.][]{Lu2020,Goodwin2023b}, or the unbound debris stream \citep[e.g.][]{Krolik2016,Spaulding2022}. Recent radio observations of TDEs are illuminating a population of events that produce prompt radio-emitting outflows well described by synchrotron emission from material ejected at the time of the stellar disruption impacting the circumnuclear medium \citep{Alexander2016,Stein2021,Goodwin2022,Goodwin2023a,Goodwin2023b}. A number of TDEs have shown a late-time radio flare up to 1000s of days after the initial optical flare, thought to be due to the late launching of a mildly relativistic jet \citep{Horesh2021,Cendes2022,Cendes2023}.

AGN also exhibit variability across the electromagnetic spectrum, including multiwavelength flares \citep{Hovatta2008,Farrar2009} thought to be due to enhancements in accretion onto the SMBH \citep{Ciotti2007}, magnetic energy release \citep{deVries1992}, disk instabilities \citep{Lightman1974,Sniegowska2020}, and at radio frequencies shocks in the existing radio jet \citep{Marscher1985}. Apparent AGN variability at radio frequencies may also be associated with interstellar scintillation or jet evolution \citep{Ross2021}. The link between variable optical and radio emission in AGN flares is not well-studied. Two scenarios exist for producing an optical flare during an AGN radio flare; the first due to a sudden increase in the mass being accreted, which could precede a radio flare by up to years depending on how and where the shock develops \citep{Pyatunina2007}. Alternatively, an optical flare may be observed directly related to a shock in the radio jet, as the optically thin tail of the shock spectrum may be detectable up to optical frequencies \citep{Valtaoja1992}. Flaring AGN are usually observed to show continuous variability over decades of monitoring \citep{Hovatta2008}.

In this work we present the discovery of a large amplitude radio flare associated with the nuclear transient event \erasst{} (hereafter J2344).
J2344 was first discovered on 2020 November 28 by the eROSITA instrument \citep{Predehl2021} on-board the SRG observatory \citep{Sunyaev2021} as a bright (0.2--2~keV $\log (L_{\mathrm{X}})\sim 44.7$), transient, ultra-soft X-ray source in the second eROSITA all-sky survey \citep{Homan2023}. The nucleus of the galaxy WISEA J234402.95-352641.8, at a redshift of $z=0.1$, brightened by a factor of at least 150 in the 0.2--2~keV X-ray band, 3\,mag in optical, and 0.3\,mag in the infra-red, with the first X-ray detection occurring $\sim$20\,d after the optical peak \citep{Homan2023}. Follow-up optical spectra taken within weeks of the X-ray and optical flare show a blue continuum, broad Balmer emission lines and narrow [OIII] and [NII] emission lines. \citet{Homan2023} analysed the early-time X-ray and optical characteristics of this transient event and found the most likely explanation for the transient emission to be a TDE within a turned-off AGN. However, they could not rule out a rapid increase in accretion in an AGN as an explanation, given the available observations. 

Here we present detailed, multi-epoch radio observations of J2344, in which we discovered transient radio emission associated with the event. In Section \ref{sec:obs} we describe the observations and data reduction. In Section \ref{sec:results} we present the results and detailed spectral and equipartition modelling of the outflow properties. In Section \ref{sec:jetoutflow} we describe the outflow properties and how they give insight into the nature of the outflow that was observed. In Section \ref{sec:discussion} we discuss the implications of the results and compare the observed properties of the transient with TDEs and AGN. Finally, in Section \ref{sec:conclusion} we summarise the results and provide concluding remarks. 

\section{Observations}\label{sec:obs}

We observed the coordinates of J2344 on 12 occasions with the Australia Telescope Compact Array (ATCA) in the frequency range 2--24\,GHz between 2021 April and 2023 June. 
All observations were taken with the ATCA CABB in the full 2048 spectral channel mode. At each epoch we observed using some combination of dual receivers with central frequencies of 2.1\,GHz, 5.5/9\,GHz, and 16.7/21.2\,GHz, which all have 2\,GHz of bandwidth. A summary of the observations is given in Table \ref{tab:radio_obs_sum}. 

All ATCA data were reduced using standard procedures in the Common Astronomy Software Application (CASA v 5.6.3; \citealt{casa2022}), including flux and bandpass calibration with PKS 1934-638 and phase calibration with PKS 2337-334. We imaged the target field using the CASA task \texttt{tclean}. For the 16.7/21.2\,GHz observations we used a cellsize of 0.12 and 0.1 arcsec and image size of 2048 pixels, for 5.5/9\,GHz we used a cellsize of 0.3 and 0.2 arcsec and image size of 3000 pixels, and for 2.1\,GHz we used a cellsize of 1 arcsec and image size of 7000 pixels, which resulted in approximately 5 pixels across the synthesized beam for each image. Larger image sizes were required at the lower frequencies in order to deconvolve bright sources in the field of view. Where enough bandwidth was available, we split each frequency band into two sub-bands for imaging, although note that occasionally at 2.1\,GHz severe RFI resulted in much of the band being flagged and only one image was created for the entire frequency band.

In each observation at each frequency we detected a point source at the location of J2344, which varied in flux density between epochs. We extracted the flux density of the target using the CASA task \texttt{imfit}, by fitting an elliptical Gaussian fixed to the size of the synthesized beam. The extracted flux densities and errors are reported in Appendix \ref{sec:radiomeasurements}. We note that the data collected on 2021-08-21 in H214 configuration were unusable due to the compact array configuration, causing J2344 to be confused with nearby bright sources in the field, so were excluded from the analysis. The 2.1\,GHz data collected on 2021-09-18 were combined with the data from 2021-10-05 to increase the sensitivity, resulting in a total of 10 spectral epochs. 

\begin{table}
	\centering
	\caption{Summary of ATCA observations of \erasst.}
	\label{tab:radio_obs_sum}
 \begin{tabular}{p{1.5cm}p{1.5cm}p{3cm}p{1.5cm}}
        \hline
        Date & Date (MJD) & Frequency (GHz) & Array config.\\
        \hline
2021-04-05	&59309	&5.5, 9	&6D\\
2021-06-06	&59371	&2.1, 5.5, 9&	6B\\
2021-08-21	&59447	&2.1, 5.5, 9&H214\\
2021-09-18	&59475	&2.1 & 6A\\
2021-10-05	&59492	&2.1, 5.5, 9, 16.7, 21.2 &6A\\
2021-11-23	&59541	&2.1, 5.5, 9, 16.7, 21.2 &6C\\
2022-01-21	&59600	&2.1, 5.5, 9, 16.7, 21.2 &6A\\
2022-04-04	&59673	&2.1, 5.5, 9, 16.7, 21.2 &6A\\
2022-08-27	&59818	&2.1, 5.5, 9, 16.7, 21.2 &6D\\
2022-12-11	&59924	&2.1, 5.5, 9, 16.7, 21.2 &6C\\
2023-03-17 & 60020 & 2.1, 5.5, 9, 16.7, 21.2 & 750C \\
2023-06-07 & 60101 & 2.1, 5.5, 9, 16.7, 21.2 & 6D \\
\hline
	\end{tabular}
\end{table}

\subsection{Archival observations}
The Rapid ASKAP Continuum Survey \citep[RACS;][]{RACs2020,RACs2021} observed the location of J2344 on 2019 Apr 29 (0.88\,GHz) and 2021 Jan 10 (1.37\,GHz), approximately 2--6 months before and after the X-ray/optical flare. There is no radio source detected in either the RACS-low (0.88\,GHz) or RACS-mid (1.37\,GHz) observations, with a 3$\sigma$ upper limit of $<0.64$\,mJy at 0.88\,GHz and $<0.63$\,mJy at 1.36\,GHz. These radio flux densities provide an upper limit for the host emission as well as the early time radio emission at $1.37$\,GHz, and indicate that there is no strong radio AGN activity in the host galaxy. We note that we cannot rule out low-luminosity AGN emission in the host, and indeed the optical spectrum of the galaxy shows strong narrow [OIII] emission lines \citep{Homan2023}; a signature of AGN activity. 

\subsection{Interstellar scintillation}

Due to the compact nature of the radio source and its distance, the observed radio flux density is expected to be affected by interstellar scintillation (ISS). To determine the amount of variability expected due to ISS, we use the NE2001 electron density model \citep{Cordes2002} and infer that for the Galactic coordinates of J2344, the transition frequency between strong and weak regimes occurs at 7.75\,GHz and the angular size limit of the first Fresnel zone at the transition frequency is 4\,microarcsec. Next, using the \citet{Walker1998} formalism as appropriate for extra-galactic sources, we find that the radio emission for J2344 will be in the strong refractive scintillation regime until the source reaches an angular size of 70\,microarcsec at 2.1\,GHz, and 8.5\,microarcsec at 5.5\,GHz. Given the distance of J2344 ($D_A\approx380$\,Mpc), the source will be in the strong scintillation regime until a source size of $\approx2\times10^{17}$\,cm at 2.1\,GHz and $\approx2.4\times10^{16}$\,cm at 5.5\,GHz, with variability expected to have a modulation fraction of 0.45--0.01 and occur on timescales of 35--2\,hrs at 2--21\,GHz.

In order to account for the variability induced by ISS between epochs at different frequencies, we introduced an additional uncertainty for each flux density measurement before carrying out the spectral fitting, added in quadrature with the statistical flux density uncertainty. This additional uncertainty varied from 45$\%$ at 2\,GHz, to 30$\%$ at 5.5\,GHz, and 10$\%$ at 9\,GHz (in the weak regime). Due to the 16.7 and 21.2\,GHz observations being well above the transition frequency in the weak regime, the variability expected due to ISS is of the order of a few percent and thus is outweighed by the statistical uncertainty. Therefore we did not include any associated flux density errors at these frequencies. All additional errors due to ISS are listed in Table \ref{tab:radio_obs}. 

\section{Results}\label{sec:results}

\begin{figure*}
    \centering
    \includegraphics[width=\columnwidth]{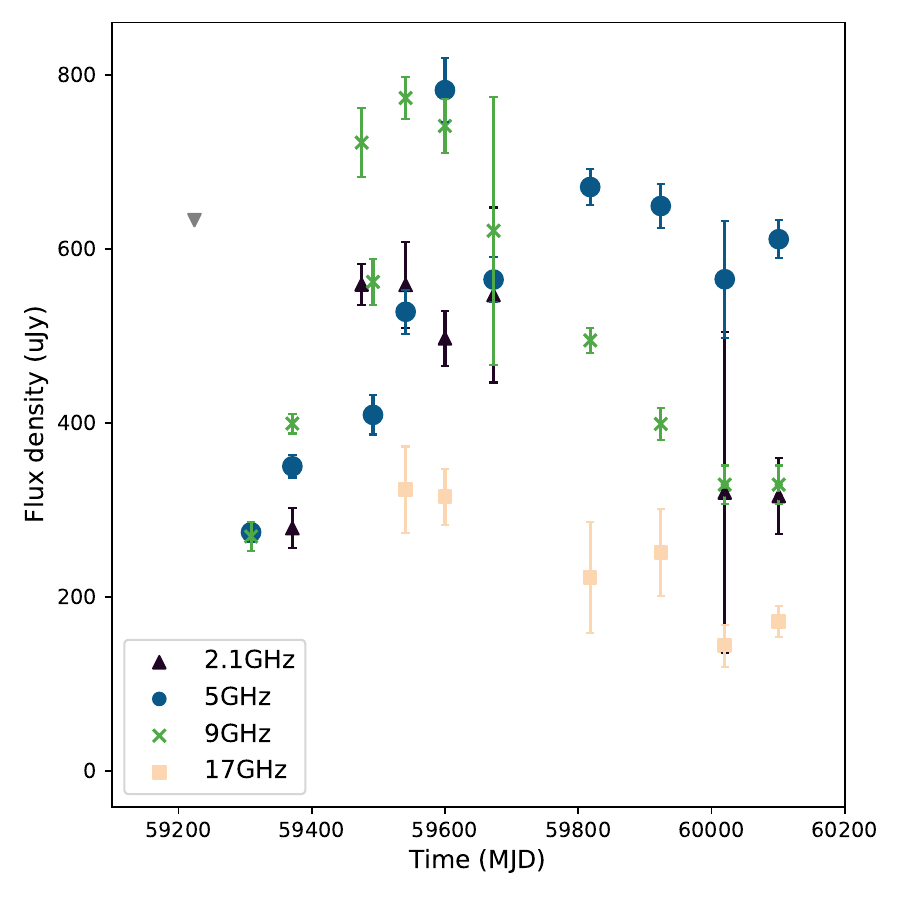}
    \includegraphics[width=\columnwidth]{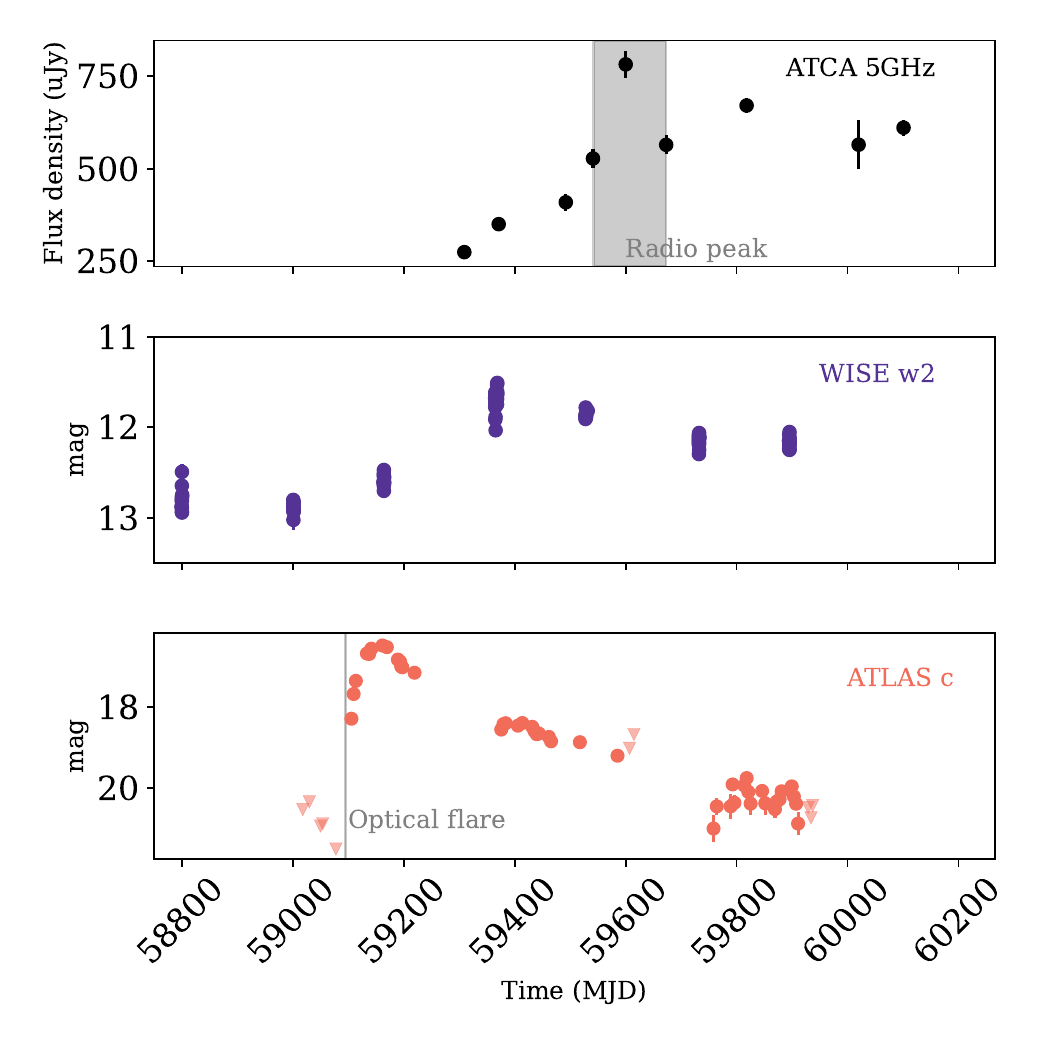}
    \caption{\textit{Left:} ATCA multi-frequency radio lightcurve of J2344. The radio spectrum peaks first at 9\,GHz (green), then 5\,GHz (blue) and lastly at 2.1\,GHz (purple), characteristic of an expanding synchrotron-emitting region. The grey inverted triangle indicates the 1.37\,GHz ASKAP-RACs 3$\sigma$ upper limit from 2021 Jan 10. \textit{Right:} ATCA 5\,GHz radio lightcurve (top), WISE w2 infra-red lightcurve (middle), and ATLAS $c$-band optical lightcurve \citep[bottom,][]{Homan2023}. The ATLAS data are extracted from difference imaging. The timing of the radio peak and beginning of the optical rise are highlighted in grey shaded regions, where the width of the region corresponds to the uncertainty in the time. The optical flare clearly leads the infra-red and radio flares, respectively. }
    \label{fig:radio_fluxes}
\end{figure*}

The radio emission associated with J2344 slowly rose to a peak at 2.1, 5.5, and 9\,GHz during the first 300\,d of radio observations (corresponding to up to 540\,d post onset of the optical flare). We observed the light curves rise and then decay at all of 9\,GHz, 5.5\,GHz, and 2.1\,GHz whereas the radio emission at >16.7\,GHz was observed to only decay throughout our monitoring (as shown in Figure~\ref{fig:radio_fluxes}). This multi-frequency evolution is characteristic of an expanding synchrotron emitting region, where the emission peaks first at higher frequencies and the peak gradually shifts to lower frequencies as the emitting region grows \citep[e.g.][]{Chevalier1998,Granot2002,Perez2001}. The time of the radio peak also trails the peak observed at infra-red and optical (with the optical occurring first, Figure \ref{fig:radio_fluxes}). This light curve behaviour is similar to other TDEs in which the radio emission was observed to rise slowly hundreds of days after the initial optical flare \citep[e.g. AT2019azh, AT2019dsg AT2020opy;][]{Stein2021,Cendes2021,Goodwin2022,Goodwin2023a}. 

\subsection{Spectral modelling}

The broadband radio spectra of J2344 are well-described by a peaked synchrotron spectrum that evolves over time. In order to constrain the synchrotron properties of the source, we fit each spectrum using the spectral fitting model described in \citet{Granot2002}, similar to the approach used in \citet{Alexander2016,Goodwin2022,Goodwin2023a,Goodwin2023b,Cendes2021}. We assume that the synchrotron emission is in the regime where the peak is associated with synchrotron self-absorption, i.e. $\nu_{\rm m} < \nu_{\rm a} < \nu_{\rm c}$ (where $\nu_{\rm m}$ is the synchrotron minimum frequency, $\nu_{\rm a}$ is the synchrotron self-absorption frequency, and $\nu_{\rm c}$ is the synchrotron cooling frequency). This is generally the case for non-relativistic outflows in which a blastwave accelerates the ambient electrons into a power law distribution $N(\gamma)\propto \gamma^{-p}$, where $\gamma$ is the electron Lorentz factor, $p$ is the synchrotron energy index, and $N$ is the density of electrons \citep[e.g.][]{BarniolDuran2013}. The synchrotron emission is then described by \citet{Granot2002} as
\begin{equation}\label{eq:F_synch}
\begin{aligned}
    F_{\nu, \mathrm{synch}} = F_{\nu,\mathrm{ext}} \left[\left(\frac{\nu}{\nu_{\rm m}}\right)^2 \exp(-s_1\left(\frac{\nu}{\nu_{\rm m}}\right)^{2/3}) + \left(\frac{\nu}{\nu_{\rm m}}\right)^{5/2}\right] \times \\
    \left[1 + \left(\frac{\nu}{\nu_{\rm a}}\right)^{s_2(\beta_1 - \beta_2)}\right]^{-1/s_2},
    \end{aligned}
\end{equation}
where $\nu$ is the frequency, $F_{\nu,\mathrm{ext}}$ is the normalisation, $s_1 = 3.63p-1.60$, $s_2 = 1.25-0.18p$, $\beta_1 = \frac{5}{2}$, and $\beta_2 = \frac{1-p}{2}$.

Due to the narrow, high ionisation emission lines of the host galaxy optical spectrum \citep{Homan2023} and the lack of a strong archival upper limit on the host radio emission, we also include a host component to the observed radio emission, such that the total observed radio emission is described by

\begin{equation}\label{eq:spectralfit}
    F_{\nu,\mathrm{total}} = F_{\nu, \mathrm{host}} + F_{\nu, \mathrm{synch}}.
\end{equation}

In order to constrain $F_{\nu, \mathrm{host}}$ and account for uncertainty in this parameter, we fit all 10 epochs simultaneously, with Equation \ref{eq:spectralfit}, where $F_{\nu, \mathrm{synch}}$ is given by Equation \ref{eq:F_synch} and the host emission is described by

\begin{equation}\label{eq:hostflux}
    F_{\nu, \mathrm{host}} = F_{0} \left( \frac{\nu}{1.4\,\mathrm{GHz}} \right)^{\alpha_0},
\end{equation}
where $F_0$ is the flux density measured at 1.4\,GHz ($F_0<0.4$\,mJy) and $\alpha_0$ is the spectral index of the host galaxy. 

We use a Python implementation of Markov Chain Monte Carlo (MCMC), \texttt{emcee} \citep{emcee} including a Gaussian likelihood function where the variance is underestimated by some fractional amount $f$. We allow $F_{\nu,\rm{ext}}$, $\nu_m$, and $\nu_a$ to be fit individually for each epoch, but only fit a single $p$, $F_0$, and $\alpha_0$ for all epochs. 
We use flat prior distributions for all parameters. The prior ranges set for each parameter are:
$10^{-6} < F_{\nu,\rm{ext}} < 10$; $0.5 < \nu_m< \nu_a$; $\nu_m< \nu_a < 10$; $2 < p < 3.5$; $0.001 < F_{0} < 0.3$; and $-0.7 < \alpha_0 < -0.3$.

Using this approach, we constrained $F_0=0.19\pm0.07$, $\alpha_0=-0.55\pm0.13$, and $p=3.41\pm0.06$. We note that $p>3$ is quite high for synchrotron emission, but is not impossible and is consistent with spectral indices of other TDE radio outflows observed \citep[e.g.][]{Goodwin2022}.

\begin{figure}
    \centering
    \includegraphics[width=\columnwidth]{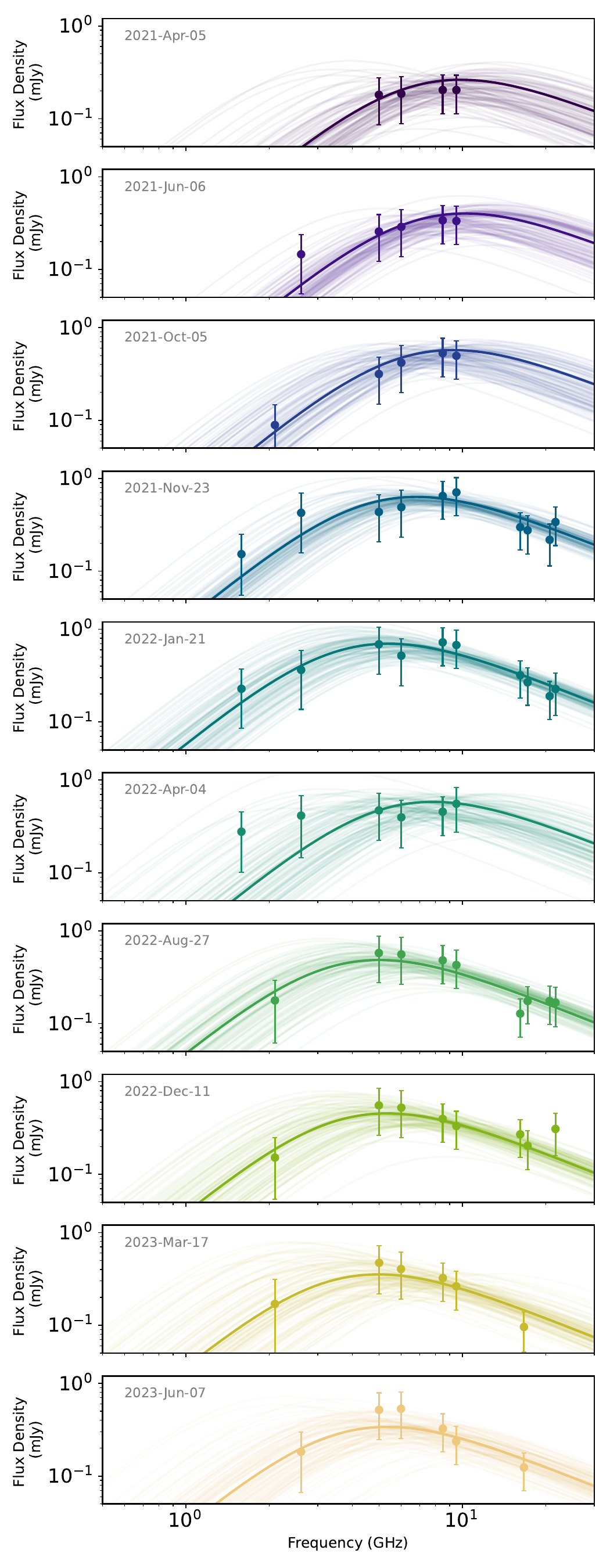}
    \caption{Spectral fits of the transient component of the radio emission assuming synchrotron emission where the peak is associated with the self-absorption break for each of the 10 epochs of ATCA observations of J2344. Solid dark lines indicate the best-fit spectrum while lighter lines indicate 100 random samples from the MCMC distribution of spectral fits to demonstrate the approximate uncertainty in the fits. It is apparent that the peak of the spectrum shifts to lower frequency over time.}
    \label{fig:specfits}
\end{figure}

\begin{figure}
    \centering
    \includegraphics[width=\columnwidth]{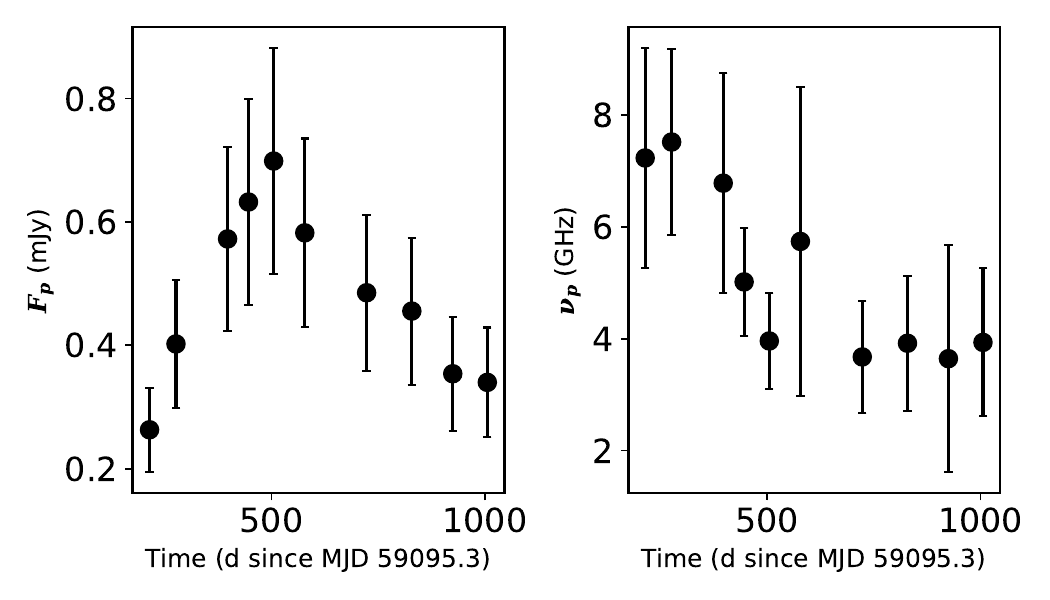}
    \caption{\textit{Left:} Peak flux density evolution with time for the synchrotron spectral fits of each epoch. \textit{Right:} Peak frequency evolution with time for the synchrotron spectral fits of each epoch. The peak flux density of the synchrotron spectrum rose for the first 300--400\,d before beginning to fade, and the peak frequency in general decreased over the course of the observations, although there was a sharp drop between 397-446\,d. }
    \label{fig:peakspec}
\end{figure}

The resulting synchrotron spectral fits for each epoch are plotted in Figure \ref{fig:specfits}, where only the transient emission is shown and the observed data points have had the constrained host component subtracted. The evolution of the peak frequency and peak flux density is plotted in Figure \ref{fig:peakspec}. The peak flux density of the synchrotron spectrum rose for the first 300--400\,d before beginning to fade, and the peak frequency in general decreased over the course of the observations, with a large drop between 397--446\,d. 

\subsection{Outflow modelling} \label{sec:outflowmodelling}

Under the assumption that the radio emission observed is described by an expanding synchrotron-emitting region, the physical outflow properties may be estimated under the assumption of equipartition. We use the synchrotron emission equipartition model from \citet{BarniolDuran2013}. In order to derive the equipartition radius, $R$ and energy, $E$, we assume equipartition between the electron and magnetic field energy densities. The exact equations we use to calculate the radius, $R$, energy, $E$, ambient electron density, $n_e$, velocity, $\beta$, and magnetic field strength, $B$ are given in \citet{Goodwin2022}. The inferred outflow properties are heavily dependent on the assumed geometry of the outflow, where the geometry is described by geometric factors given by $f_A = A/(\pi R^2 / \Gamma^2)$ and $f_V = V/(\pi R^3 / \Gamma^4)$, for area, $A$, and volume, $V$, of the outflow, and distance from the origin of the outflow, $R$ \citep{BarniolDuran2013}. Since the outflow geometry is not known, we include two geometries in our analysis: a "spherical" geometry where $f_A=1$ and $f_V=4/3$; and a "conical" geometry appropriate for a mildly collimated jet where $f_A=0.13$ and $f_V=1.15$. 

Assuming equipartition allows us to estimate the key physical quantities, however, the emitting region is likely not in equipartition. We therefore apply a post-correction to the physical quantities to correct for any expected deviation from equipartition. For this correction, we assume that the fraction of the total energy of the outflow in the magnetic field is 2$\%$, i.e. $\epsilon_B=0.02$ based on observations of TDEs and supernovae \citep{Horesh2013,Cendes2021b}. We additionally assume that 10$\%$ of the total energy is carried by the electrons, i.e. $\epsilon_e=0.1$ as has been adopted in other TDE outflow analyses \citep[e.g.][]{Alexander2016,Cendes2021b,Goodwin2022}. However, we note that recent studies have found $\epsilon_e\sim10^{-3}-10^{-4}$ for non-relativistic collisionless shocks \citep{Park2015,Xu2020}. For reference, in Appendix \ref{sec:equideviations} we include the outflow properties calculated for different assumed $\epsilon_e$ and note that smaller $\epsilon_e$ results in slightly larger radii and energies and a lower inferred ambient electron density.

\begin{figure*}
    \centering
    \includegraphics[width=2\columnwidth]{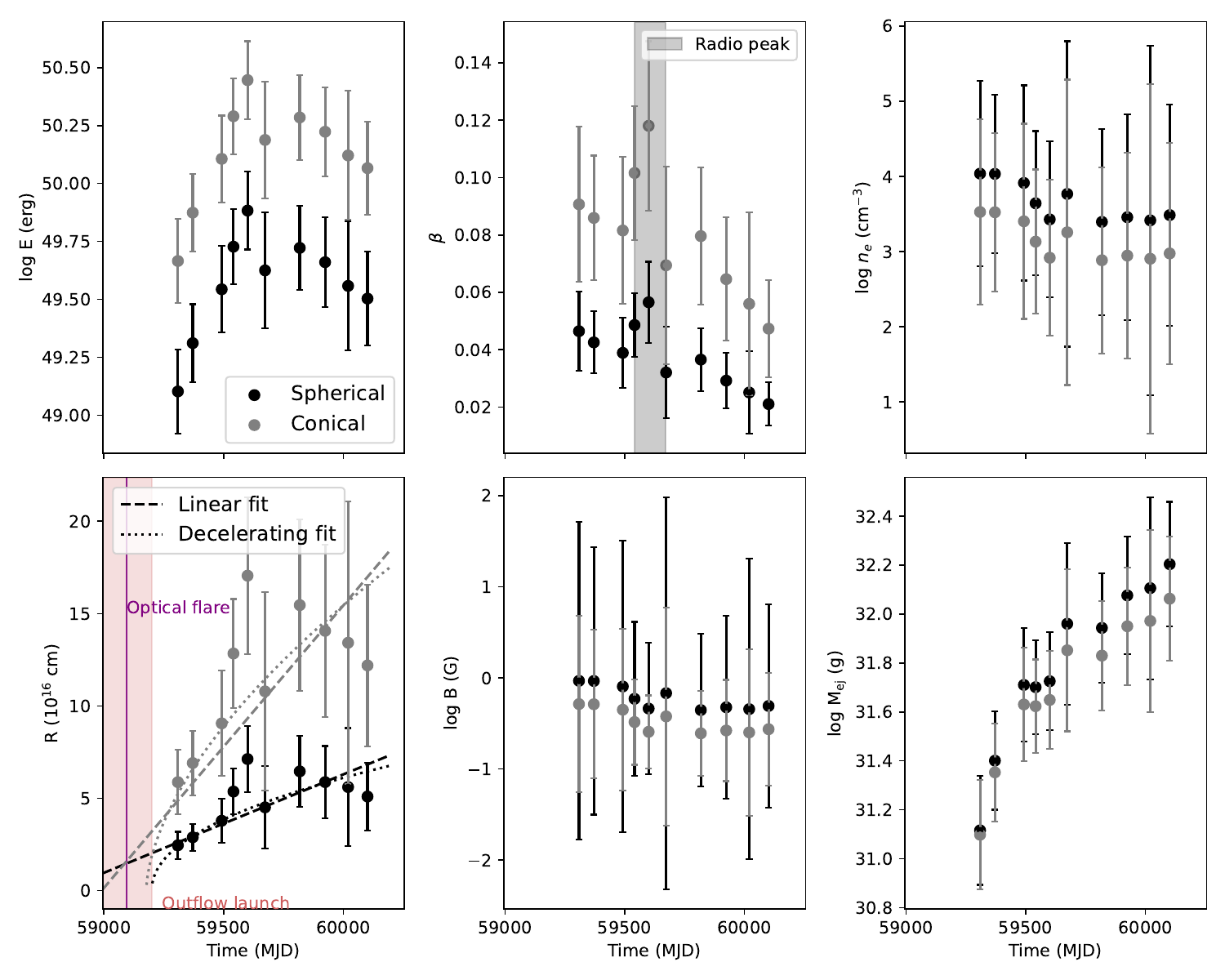}
    \caption{Physical outflow properties inferred from equipartition modelling of the spectral properties of the radio emission from J2344.
    Properties derived with an assumed spherical geometry are plotted in black, and ones with a conical geometry in grey. $E$ and $R$ are the estimated energy and radius of the outflow derived from an equipartition analysis and corrected for assumed deviation from equipartition. $\beta$ is the outflow velocity divided by the speed of light, $B$ is the magnetic field strength, $n_e$ is the free electron number density of the ambient medium, and $M_{\rm ej}$ is the mass in the ejecta. The dashed lines in the lower left panel show a linear fit to the radius for each geometry and the dotted lines show a power-law fit to the radius for each geometry assuming a decelerating outflow. The onset of the optical rise time is indicated in purple shading and the estimated outflow launch date (when $R=0$) is indicated in red shading.}
    \label{fig:outflowprops}
\end{figure*}

The calculated properties of the outflow are listed in Table \ref{tab:outflowprops} and plotted in Figure \ref{fig:outflowprops}. 

\begin{table*}
    \centering
    \caption{Equipartition modelling properties of the outflow produced by J2344 based on the synchrotron spectral fits. We report both the uncorrected equipartition radius ($R_{\rm{eq}}$) and energy ($E_{\rm{eq}}$) as well as the corrected radius ($R$) and energy ($E$). The time for each epoch is measured with respect to the optical flare date, MJD 59059}
    \begin{tabular}{p{0.7cm}cp{1cm}ccccccccc}
&Time & $F_{\rm{p}}$ & $\nu_{\rm{p}}$  & log $R_{\rm{eq}}$  & log $E_{\rm{eq}}$ & log $R$  & log $E$  & $\beta$ & log $B$ & log $n_e$  & log $M_{\rm{ej}}$ \\

&(d) & (mJy) &  (GHz) & (cm) & (erg) & (cm) & (erg) &  & (G) & (cm$^{-3}$) & (g) \\

\hline
\hline
&214 & 0.26$\pm$0.07 & 7.23$\pm$1.96       & 16.42$\pm$0.13&  49.06$\pm$0.18 & 16.39$\pm$0.13 &       49.10$\pm$0.18 & 0.05$\pm$0.01 &       -0.03$\pm$1.75      & 4.04$\pm$1.23      & 31.12$\pm$0.22 \\
&276 & 0.40$\pm$0.10 & 7.52$\pm$1.66       & 16.49$\pm$0.11&  49.27$\pm$0.17 & 16.46$\pm$0.11 &       49.31$\pm$0.17 & 0.04$\pm$0.01 &       -0.03$\pm$1.47      & 4.03$\pm$1.05      & 31.40$\pm$0.20 \\
&397 & 0.57$\pm$0.15 & 6.78$\pm$1.96       & 16.60$\pm$0.14&  49.50$\pm$0.19 & 16.58$\pm$0.14 &       49.54$\pm$0.19 & 0.04$\pm$0.01 &       -0.09$\pm$1.60      & 3.91$\pm$1.30      & 31.71$\pm$0.23 \\
&446 & 0.63$\pm$0.17 & 5.02$\pm$0.97       & 16.75$\pm$0.10&  49.68$\pm$0.16 & 16.73$\pm$0.10 &       49.73$\pm$0.16 & 0.05$\pm$0.01 &       -0.23$\pm$0.84      & 3.64$\pm$0.96      & 31.70$\pm$0.19 \\
Spherical&505 & 0.70$\pm$0.18 & 3.96$\pm$0.86       & 16.88$\pm$0.11&  49.84$\pm$0.17 & 16.85$\pm$0.11 &       49.88$\pm$0.17 & 0.06$\pm$0.01 &       -0.34$\pm$0.72      & 3.43$\pm$1.04      & 31.73$\pm$0.20 \\
&578 & 0.58$\pm$0.15 & 5.74$\pm$2.76       & 16.68$\pm$0.22&  49.58$\pm$0.25 & 16.65$\pm$0.22 &       49.62$\pm$0.25 & 0.03$\pm$0.02 &       -0.17$\pm$2.15      & 3.77$\pm$2.03      & 31.96$\pm$0.33 \\
&723 & 0.49$\pm$0.13 & 3.67$\pm$1.00       & 16.84$\pm$0.13&  49.68$\pm$0.18 & 16.81$\pm$0.13 &       49.72$\pm$0.18 & 0.04$\pm$0.01 &       -0.35$\pm$0.84      & 3.40$\pm$1.24      & 31.94$\pm$0.22 \\
&829 & 0.46$\pm$0.12 & 3.92$\pm$1.21       & 16.79$\pm$0.14&  49.62$\pm$0.19 & 16.77$\pm$0.14 &       49.66$\pm$0.19 & 0.03$\pm$0.01 &       -0.32$\pm$1.00      & 3.46$\pm$1.37      & 32.08$\pm$0.24 \\
&925 & 0.35$\pm$0.09 & 3.64$\pm$2.03       & 16.77$\pm$0.25&  49.51$\pm$0.28 & 16.75$\pm$0.25 &       49.56$\pm$0.28 & 0.03$\pm$0.01 &       -0.34$\pm$1.65      & 3.42$\pm$2.33      & 32.11$\pm$0.37 \\
&1006 & 0.34$\pm$0.09 & 3.93$\pm$1.32       & 16.73$\pm$0.16&  49.46$\pm$0.20 & 16.71$\pm$0.16 &       49.50$\pm$0.20 & 0.02$\pm$0.01 &       -0.31$\pm$1.12      & 3.49$\pm$1.48      & 32.20$\pm$0.25 \\
\hline
&214 & 0.26$\pm$0.07 & 7.23$\pm$1.96       & 16.79$\pm$0.13&  49.62$\pm$0.18 & 16.77$\pm$0.13 &       49.67$\pm$0.18 & 0.09$\pm$0.03 &       -0.29$\pm$0.97      & 3.53$\pm$1.23      & 31.10$\pm$0.22 \\
&276 & 0.40$\pm$0.10 & 7.52$\pm$1.66       & 16.87$\pm$0.11&  49.83$\pm$0.17 & 16.84$\pm$0.11 &       49.87$\pm$0.17 & 0.09$\pm$0.02 &       -0.29$\pm$0.82      & 3.52$\pm$1.05      & 31.35$\pm$0.20 \\
&397 & 0.57$\pm$0.15 & 6.78$\pm$1.96       & 16.98$\pm$0.14&  50.06$\pm$0.19 & 16.96$\pm$0.14 &       50.11$\pm$0.19 & 0.08$\pm$0.03 &       -0.35$\pm$0.89      & 3.40$\pm$1.30      & 31.63$\pm$0.23 \\
&446 & 0.63$\pm$0.17 & 5.02$\pm$0.97       & 17.13$\pm$0.10&  50.25$\pm$0.16 & 17.11$\pm$0.10 &       50.29$\pm$0.16 & 0.10$\pm$0.02 &       -0.48$\pm$0.47      & 3.13$\pm$0.96      & 31.62$\pm$0.19 \\
Conical&505 & 0.70$\pm$0.18 & 3.96$\pm$0.86       & 17.26$\pm$0.11&  50.40$\pm$0.17 & 17.23$\pm$0.11 &       50.45$\pm$0.17 & 0.12$\pm$0.03 &       -0.59$\pm$0.40      & 2.92$\pm$1.04      & 31.65$\pm$0.20 \\
&578 & 0.58$\pm$0.15 & 5.74$\pm$2.76       & 17.06$\pm$0.22&  50.14$\pm$0.25 & 17.03$\pm$0.22 &       50.19$\pm$0.25 & 0.07$\pm$0.03 &       -0.42$\pm$1.20      & 3.26$\pm$2.03      & 31.85$\pm$0.33 \\
&723 & 0.49$\pm$0.13 & 3.67$\pm$1.00       & 17.21$\pm$0.13&  50.24$\pm$0.18 & 17.19$\pm$0.13 &       50.28$\pm$0.18 & 0.08$\pm$0.02 &       -0.61$\pm$0.47      & 2.89$\pm$1.24      & 31.83$\pm$0.22 \\
&829 & 0.46$\pm$0.12 & 3.92$\pm$1.21       & 17.17$\pm$0.14&  50.18$\pm$0.19 & 17.15$\pm$0.14 &       50.22$\pm$0.19 & 0.06$\pm$0.02 &       -0.58$\pm$0.56      & 2.95$\pm$1.37      & 31.95$\pm$0.24 \\
&925 & 0.35$\pm$0.09 & 3.64$\pm$2.03       & 17.15$\pm$0.25&  50.08$\pm$0.28 & 17.13$\pm$0.25 &       50.12$\pm$0.28 & 0.06$\pm$0.03 &       -0.60$\pm$0.92      & 2.91$\pm$2.33      & 31.97$\pm$0.37 \\
&1006 & 0.34$\pm$0.09 & 3.93$\pm$1.32       & 17.11$\pm$0.16&  50.02$\pm$0.20 & 17.09$\pm$0.16 &       50.07$\pm$0.20 & 0.05$\pm$0.02 &       -0.56$\pm$0.62      & 2.98$\pm$1.48      & 32.06$\pm$0.25 \\
\hline


    \end{tabular}
    \label{tab:outflowprops}
\end{table*}

\section{Constraints on the outflow launched}\label{sec:jetoutflow}

The radius constraints obtained via the equipartition modelling enable a launch date of the outflow to be estimated. A simple linear fit to the radii assuming constant velocity of the outflow results in an outflow launch date of MJD 58784$\pm3$\,d (spherical) or MJD 58953$\pm3$\,d (conical), earlier than the estimated onset of the optical flare date \citep[MJD 59095$\pm$1,][]{Homan2023}. The reduced-$\chi^2$ statistic for these two fits are 5.39 and 6.7 respectively, indicating the data are poorly fit by a single linear model. We instead fit the first 5 epochs with a linear model, before the onset of deceleration of the outflow. In these cases we constrain an outflow launch date of MJD 59054$\pm1$\,d (spherical) or MJD 59041$\pm3$\,d (conical), just 5--18\,d earlier than the estimated onset of the optical flare date. 

Alternatively, assuming the outflow velocity is not constant and the outflow is decelerating with time, we also fit a power-law to the predicted radius evolution with time, where the radius evolution with time is described by:
\begin{equation}
    R = A(t - t_0)^{\alpha}
\end{equation}
where $t_0$ is the outflow launch time. Using this equation we constrain $\alpha=0.47^{+0.07}_{-0.08}$, $t_0=59164\pm60$\,MJD (spherical) or $\alpha=0.59^{+0.06}_{-0.06}$, $t_0=59141\pm55$\,MJD (conical). The reduced-$\chi^2$ statistic for these two fits are 4.17 and 4.49 respectively, indicating a better fit than the simple linear model above. In the case of a decelerating outflow, the outflow launch date is therefore approximately 27--165\,d after the observed onset of the optical flare.

While the case of a decelerating outflow predicts an outflow launch date $t_0$ approximately consistent with or up to 165\,d after the onset of the optical rise date (Fig. \ref{fig:outflowprops} lower-left panel), a linear fit to the radius assuming constant velocity for the entire span of the observations predicts an outflow launched up to 200\,d prior to the optical flare, or only for the first 5 epochs predicts an outflow launched up to 20\,d prior to the optical flare. This modelling indicates that a decelerating model is supported if the outflow was launched around the time of the optical flare.

\subsection{Single ejection outflow}

The increasing energy during the first $\approx$500\,d post-optical flare of the outflow could be either due to an off-axis relativistic jet decelerating and widening into view (which we explore in Section \ref{sec:offaxisjet}), or a freely coasting shock front launched at the time of the optical/X-ray/UV flare. In this latter scenario, a single ejection of material creates a shock front that interacts with the surrounding material, the CNM, similar to the forward shock in a supernova \citep{Weiler1986,Chevalier1982b,Chevalier1982c,Chevalier1998,Weiler2002}. The geometry of this ejecta depends on the mechanism that produced it, where  our spherical model in Table \ref{tab:outflowprops} corresponds to an outflow launched by a TDE from stream-stream collisions or disk winds, and our conical model in Table \ref{tab:outflowprops} corresponds to a collimated outflow launched either by a TDE accretion episode or sudden accretion epsiode in a pre-existing AGN.
Regardless of the ejecta geometry, during the expansion, the shock front interacts with and heats the CNM material, sweeping up mass and transferring energy to the shocked region. This scenario has also been suggested to explain the increasing energy observed for the TDE AT2019dsg \citep{Matsumoto2022}. The electron cooling timescale is much longer than the dynamical time, where the cooling timescale is related to the magnetic field via

\begin{equation}
    t_{\rm{cool}} = 4500\,\rm{d} \left(\frac{B}{1G}\right)^{-2}\left(\frac{\gamma_m}{2}\right)^{-1}.
\end{equation}

Therefore negligible energy is lost to cooling of the electron population over the course of our observations. 

\citet{Matsumoto2022} showed that for a freely expanding outflow that is increasing in energy by sweeping up material from the CNM, the expected evolution of kinetic energy with time is dependent on the gradient of the density of the CNM, $k$, such that
\begin{equation}
    E\propto t^{(5-k)\alpha -2},
\end{equation}
where $\alpha$ is the power-law index, $R\propto(\Delta t - t_0)^{\alpha}$. In Section \ref{sec:deceleration}, we constrain for a decelerating outflow $\alpha=0.47^{+0.07}_{-0.08}$ (spherical), $\alpha=0.59^{+0.06}_{-0.06}$ (conical), or $\alpha=1$, assuming constant velocity and therefore linear radial growth with time. 

Fitting a power-law to the kinetic energy over all observations of J2344 in Figure \ref{fig:outflowprops}, $E \propto (t-t_0)^{\beta}$, we find $\beta=0.46^{+0.18}_{-0.11}$ (decelerating) or $\beta=0.41^{+0.06}_{-0.04}$ (constant velocity) for the spherical case, and  $\beta=0.62^{+0.17}_{-0.09}$ (decelerating) or $\beta=0.62^{+0.09}_{-0.06}$ (constant velocity) for the conical case. We therefore find $k=0.2-1$ (decelerating, spherical/conical) or $k=2.3-2.6$ (constant velocity, spherical/conical). 

Additionally, a power-law fit to the predicted radius and ambient density in Figure \ref{fig:outflowprops} gives $n\propto r^{-1.4}$, i.e. $k=1.4$, consistent with the CNM density gradients we infer from the energy modelling above. Although we note the large error bars on the ambient density estimates mean this fit is not particularly well-constrained. Therefore, the initially increasing energy/flux density of the radio flare is well explained by a single injection of material into an initially ballistically expanding outflow shocking the CNM, sweeping up material, and producing additional energy which begun decelerating after the peak radio brightness. Similar increasing energies have been observed in synchrotron-emitting outflows from supernovae and attributed to CNM interactions \citep{Salas2013,Anderson2017} and in TDEs such as AT2019dsg \citep{Cendes2021,Stein2021} and AT2019azh \citep{Goodwin2022}. 

\subsubsection{Deceleration time and peak radio flux density}\label{sec:deceleration}

The peak radio flux density for J2344 initially rose, until MJD 59600$\pm$65\,d (505\,d post-optical flare) at which time the peak flux density began to fall (Figure \ref{fig:peakspec}). When the outflow sweeps up mass approximately equal to the initial mass in the outflow, it will begin to decelerate, and no longer produce additional energy. The decay of the radio emission and transition to constant kinetic energy can be explained by the deceleration of the outflow. In the deceleration phase, the light curve is expected to evolve following a Sedov-Taylor decay.  In the case of a freely expanding outflow, theoretically, the time of peak radio flux density should correspond to the deceleration time, or the time at which the outflow has swept up equivalent mass from the CNM to the original mass in the ejecta \citep{Lu2020}. After this time, the outflow will enter the Sedov-Taylor decay-phase if no additional energy is deposited. The deceleration radius for a freely expanding shock front is given by \citet{Lu2020}
\begin{equation}
    r_{\rm{dec,pc}}^{3-k} = \frac{3-k}{\Omega}\frac{2E_k}{N_{\rm{pc}}m_pv_0^2},
\end{equation}
where $N_{\rm{pc}} = n_{\rm{pc}} (1\rm{pc})^3$ is a reference number of electrons, $m_p$ is the proton mass, $v_0$ is the initial velocity, $E_k$ is the kinetic energy of the outflow, $\Omega$ is the solid angle the outflow covers (for a spherical outflow we assume $\Omega=2\pi$), and $k$ is related to the ambient density profile via $n = n_{\rm{pc}} r_{\rm{pc}}^{-k}$ ($k<3$).

For J2344, the outflow modelling in Section \ref{sec:outflowmodelling} produces significantly different outflow energies, velocities, radii, and ambient densities depending on assumptions about the geometry and equipartition. Therefore, taking the widest range in the outflow parameters (for changing geometry and equipartition), we find for a kinetic energy of $E_k\sim1\times10^{49}$\,erg to $E_k\sim1\times10^{53}$\,erg , velocity$\sim$0.05$c$ to $\sim0.3c$, and ambient density $n_e\sim10^3$\,cm$^{-3}$ to $n_e\sim10^{-1}$\,cm$^{-3}$, and assuming $k=2$ and $\Omega=2\pi$ we find the deceleration radius could be anywhere from  8$\times10^{14}$\,cm to  2$\times10^{21}$\,cm and the onset of deceleration time could be anywhere from 6\,d to 8000\,yr post outflow launch. If we take the peak radio flux density of the lightcurve to correspond to the deceleration time, this occurred between 397--505\,d post outflow launch at a predicted radius of $5.3\times10^{16}$\,cm (spherical) or $1.3\times10^{17}$\,cm (conical). While the predicted deceleration time is subject to many assumptions and is therefore quite uncertain, the observed time of the radio peak is within the expected time range at which the outflow should have begun decelerating if the radio emission observed is described by an initially ballistic outflow model. Additionally, the deceleration time measured would imply the equipartition parameters are closer to those presented in Table \ref{tab:outflowprops} than those in Table \ref{tab:outflowprops_appendix}, i.e. the fraction of total energy of the outflow carried by the electrons is closer to 10\% than 0.01\% or lower.

In this deceleration phase, the flux density evolution of the fading shockwave is proportional to the CNM density gradient and synchrotron spectral indices \citep{Sironi2013} and can be described by a spherical shock-front impacting a stratified medium with density profile $n\propto r^{-k}$ and the electrons in the shock are accelerated into a power-law distribution $dN_e/d\gamma \propto \gamma^{-p}$, where $\gamma$ is the electron Lorentz factor. The radio luminosity is thus related to time via

\begin{equation}\label{eq:sedovL}
    L_{\nu} = \nu^{(1-p)/2} t^{-(2(3-k)(p-3)+3(p+1)/(2(5-k)))}.
\end{equation}

\begin{figure}
    \centering
    \includegraphics[width=0.8\columnwidth]{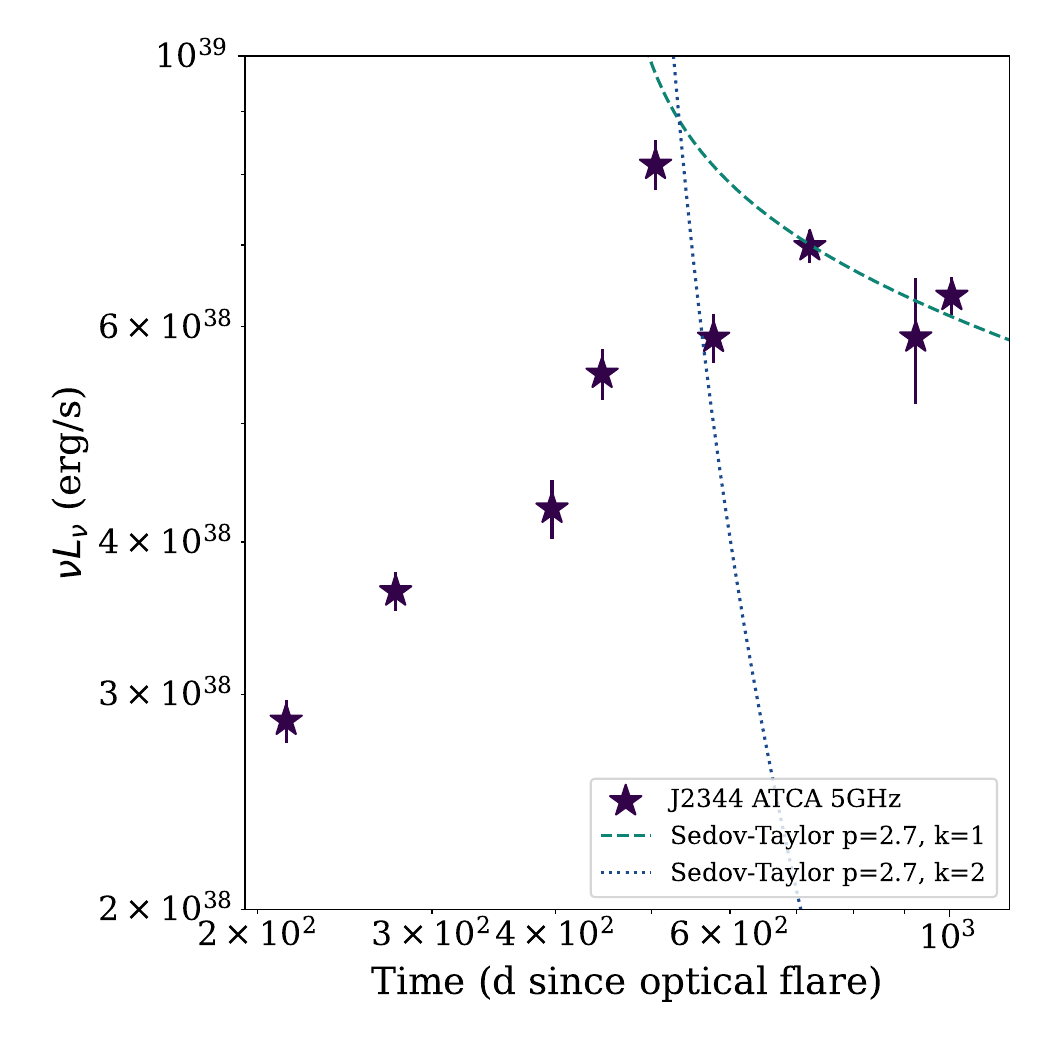}
    \caption{Radio luminosity of J2344 and the predicted flux density evolution during the Sedov-Taylor decay phase for two different CNM density distributions, $n\propto r^{-k}$, assuming an electron energy index $p=2.7$.}
    \label{fig:sedovdecay}
\end{figure}

In Figure \ref{fig:sedovdecay} we plot the 5\,GHz luminosity evolution of J2344 as well as two predicted Sedov-Taylor luminosity decays for $k=1$ and $k=2$, fixing $p=2.7$. Evidently, the flatter $k=1$ CNM density gradient is preferred for the evolution of the luminosity decay of J2344. We therefore find that both the early pre-radio peak evolution and late post-radio peak evolution of the outflow emission can be explained by a single ejection of material from near the central SMBH, and a ballistic outflow impacting a CNM with a relatively flat density profile such that $n\propto r^{-1}$. 


\begin{figure}
    \centering
    \includegraphics[width=\columnwidth]{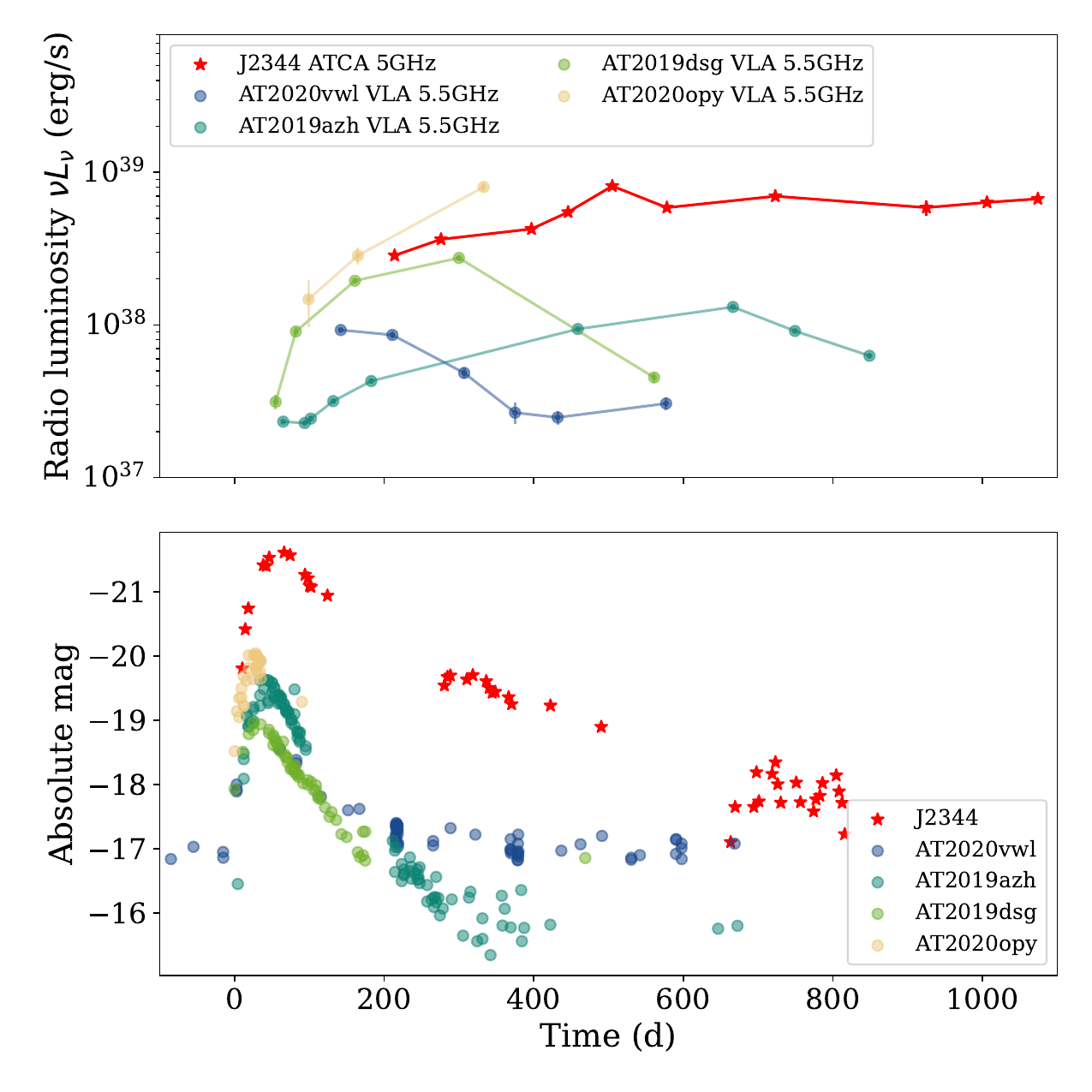}
    \caption{\textit{Top:}Radio luminosity of J2344 (red stars) and a selection of other radio-detected TDEs with good light curve coverage. \textit{Bottom:} Optical ATLAS $c$-band light curve of J2344 \citep{Homan2023} and Gaia $g$-band light curves of the same selection of TDEs, converted to absolute magnitude.  The ATLAS data are extracted from difference imaging. All Gaia lightcurves were obtained from the Gaia photometric science alerts data base \citep[\url{http://gsaweb.ast.cam.ac.uk/alerts};][]{Gaia2016}). Times on the x-axis have been scaled to the approximate onset of the optical rise for each event. The flatter radio decay of J2344 may be due to a flatter density distribution in the central regions of the host galaxy.}
    \label{fig:tde_comparison}
\end{figure}

\subsection{Jet-like outflow}\label{sec:outflowprops}

Alternatively, the observed radio flare may be a signature of a jet that was launched coincident with the optical flare, or shocks in a pre-existing jet of an AGN (i.e. our conical model in Table \ref{tab:outflowprops}), although the archival non-detections of the host galaxy in the ASKAP-RACS radio survey rule out a bright pre-existing AGN jet in the system. In this scenario, the increasing energy may be due to continuous energy injection into the outflow from the accreting SMBH. The rate of increase in energy enables a constraint on the required luminosity injection rate, i.e.
\begin{equation}
    L_{\rm{in}} \sim \frac{dE_{\rm{eq}}}{dt} \approx 3.36\times10^{43}\,\left(\frac{t}{64\rm{d}}\right)^{0.95}\,\rm{erg\,s}^{-1},
\end{equation}
where 64\,d is the time from optical flare to peak optical brightness \citep{Homan2023} and we have assumed $E_{\rm{eq}} = A t^{B}$, constraining $A=(3.07\pm5)\times10^{44}$ and $B=1.95\pm0.3$ by fitting the constrained energies during the period the energy was increasing (Fig \ref{fig:outflowprops}). 

In comparison, the fallback luminosity of a solar mass star disrupted by a $10^7$\,$M_{\odot}$ black hole is 

\begin{equation}
    L_{\rm{fb}}\sim\dot{M}c^2\sim10^{47} \left(\frac{t}{111\rm{d}}\right)^{-5/3} R_{*} M_{*}^{1/3} M_{\rm{SMBH},7}^{-2/3}\quad\rm{erg/s}
\end{equation}
where $R_{*}$ and $M_{*}$ are the radius and mass of the disrupted star respectively in solar mass and radii units, $M_{\rm{SMBH},7}$ is the mass of the black hole in $10^7$\,$M_{\rm{\odot}}$. 

The fallback luminosity is approximately four orders of magnitude larger than the luminosity required by the energy injection rate if the increasing energy were powered by injection into the outflow from the accreting SMBH. In order for the increasing energy of the outflow to be explained by continuous energy injection from accretion, the efficiency of accretion would have to be extremely high, with only 0.03\% of the accreted material being injected into the outflow. 

Additionally, the X-ray emission observed from this event also enables an estimate of the accretion rate. The observed X-ray luminosity is $L_{\rm{0.2-2\,\rm{keV}}}=7.94\times10^{44}$\,erg\,s$^{-1}$ \citep{Homan2023}. Assuming approximately 10$\%$ of the accreted mass is converted into the 0.2--2\,keV luminosity \citep[e.g.][]{Auchettl2018}, this implies an accretion luminosity of $L_{\rm{acc}}\sim8\times10^{45}$\,erg\,s$^{-1}$, meaning that just 0.8$\%$ of the accreted material is required to be injected into the outflow to power the increasing energy observed, again requiring extremely efficient accretion.

\subsubsection{Off-axis relativistic jet-like outflow}\label{sec:offaxisjet}

Recently, \citet{Matsumoto2023} extended the equipartition method to relativistic off-axis emitters. Here we assess whether the radio emission from J2344 is well-described by an off-axis relativistic jet-like outflow. In the case of an off-axis relativistic jet, the initially relativistic jet eventually decelerates and becomes Newtonian, at which time the two branches of the equipartition solutions merge and a single solution for the outflow is obtained. In the case of an initially relativistic, off-axis jet, at this transition time the equipartition Newtonian velocity must be $\beta_{\rm{Eq,N}}>0.23$, where \citep{Matsumoto2023} define

\begin{equation}
\label{eq:betaeqN}
\begin{split}
    \beta_{\rm{Eq,N}} \approx 0.73 \left[ \frac{(F_p/\rm{mJy})^{8/17} (d_L/10^{28}\rm{cm})^{16/17} \eta^{35/51}}{(\nu_p/10\rm{GHz}) (1+z)^{8/17}} \left(\frac{t}{100\,d}\right)^{-1}\right]     \\
    \times f_A^{-7/17} f_V^{-1/17}
\end{split}
\end{equation}

where all parameters are as defined in Section \ref{sec:outflowmodelling}. 

For J2344, $\beta_{\rm{Eq,N}}$ only exceeds 0.23 for jet geometries with a half-opening angle $<15$\,deg, and only for the highest velocity epochs at 440--550\,d post optical flare. For the two geometries considered throughout this work $\beta_{\rm{Eq,N}}$ never exceeds 0.23, and has been decreasing since the peak at $\approx$500\,d. 

The observations we present of J2344 indicate that both the peak flux density and peak frequency have been declining for the past 500\,d. The peak flux density is declining as $F_p\propto t^{-1.05}$ and the peak frequency has been declining as $\nu_p\propto t^{-0.35}$ (Figure \ref{fig:peakspec}). Therefore, the equipartition Newtonian velocity (Equation \ref{eq:betaeqN}) is proportional to $\beta_{\rm{Eq,N}} \propto t^{-1.14}$. Similarly to the case of AT2019dsg discussed by \citet{Matsumoto2023}, unless these trends change and either $F_p$ begins to increase or $\nu_p$ begins to decrease more rapidly for J2344, the transition from the Newtonian branch to the relativistic branch of the equipartition solutions will never happen. In this scenario, the outflow cannot be described by an off-axis relativistic jet launched by a tidal disruption event. 

\section{Discussion}\label{sec:discussion}

Our broadband multi-epoch radio spectral observations reveal a radio flare associated with the nuclear transient J2344. The radio flare rose to a peak luminosity of $\sim10^{39}$\,erg/s over $\sim500$\,d post-optical flare start and showed spectral evolution characteristic of an expanding synchrotron-emitting region due to an outflow. Through modelling of the synchrotron emission and outflow properties using an equipartition approach, we infer that the outflow was launched approximately coincident with the onset of the optical flare (within 100\,d). The question remains: what kind of nuclear transient event can explain the multi-wavelength properties that were observed?  

\citet{Homan2023} analysed the initial optical, X-ray, UV, and infra-red flare that
was observed from J2344. They deduced that the soft X-ray spectrum, rapid onset of decay in X-ray, UV, and optical, and optical spectrum are indicative of a TDE. However, they note that the high-ionisation narrow lines present in the optical spectrum are indicative that the galaxy is likely a low-luminosity AGN or was in a more active AGN phase as recently as a few millennia ago. Below we discuss the likely nature of the transient event in the context of the radio flare that we discovered.    

We modelled the synchrotron emission using an equipartition approach and found the radius increased over the 1000\,d of radio observations while the ambient density decreased, the mass in the emitting region increased, and the magnetic field strength remained approximately constant. The radio emission is consistent with an outflow with radii from $10^{16}$--$10^{17}$\,cm, energy $10^{49}$--$10^{50}$\,erg, and velocity 5--10\%\,c. Interestingly, the total kinetic energy of the outflow increased steadily until $\sim500$\,d post-outflow launch, at which time the energy plateaued. The plateau in energy corresponds to the peak radio luminosity of the light curve. At the same time, the velocity also remained approximately constant during the energy plateau phase, following which it declined.

\subsection{TDE interpretation}
TDEs are known to launch outflows that are observed as transient radio emission evolving on timescales of $\sim$months \citep[e.g.][]{Alexander2020}. The mechanism that launches non-relativistic outflows is currently under debate, with leading theories involving either a mildly collimated jet \citep[e.g.][]{Stein2021,Cendes2022}, disk-wind \citep[e.g.][]{Alexander2016}, debris stream collisions \citep{Lu2019}, or the unbound debris stream \citep{Krolik2016}. Recent studies have unveiled a population of prompt-radio emitting TDEs in which the outflow radius can be tracked back to launch dates coincident with the optical flare \citep[e.g.][]{Goodwin2023b,Goodwin2023a,Goodwin2022,Cendes2021}, suggesting prompt launching of the outflow from either debris stream collisions or the unbound debris stream. Most models predict the debris circularisation would occur on longer timescales and thus delay the production of an outflow launched via accretion processes (i.e. jet or disk winds). These types of outflows appear to also be possible, with a new population of radio-emitting TDEs recently discovered, that appear to have launched outflows 100s to 1000s of days after the initial optical flare \citep{Cendes2023}, in contrast to the prompt radio-emitters.

The inferred outflow launch date for J2344, approximately coincident with the initial optical flare, would imply an outflow launched by either debris stream collisions or the unbound debris stream. The outflow properties in this model (stream collisions spherical, unbound debris conical, Table \ref{tab:outflowprops}) are very consistent with both of those scenarios \citep[velocity $\sim0.05-0.10c$, energy $\sim10^{49}$\,erg;][]{Krolik2016,Lu2020}. Since the outflow was likely launched around the time of the initial optical flare, in this scenario it is therefore related to the optical flare that was observed from the nucleus of this galaxy. The radio observations of J2344 further enhance the case of the transient being triggered by a TDE within a low-luminosity AGN (LLAGN). 

\subsubsection{Comparison to other TDEs}

\begin{figure*}
    \centering
    \includegraphics[width=\columnwidth]{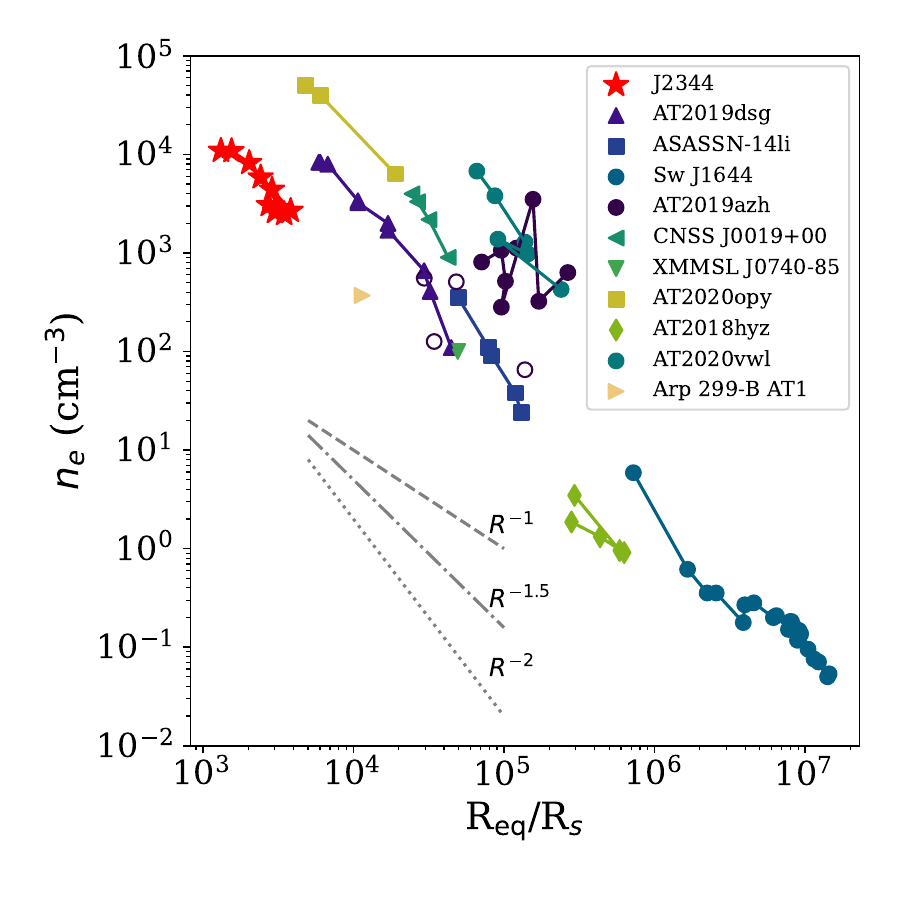}
    \includegraphics[width=\columnwidth]{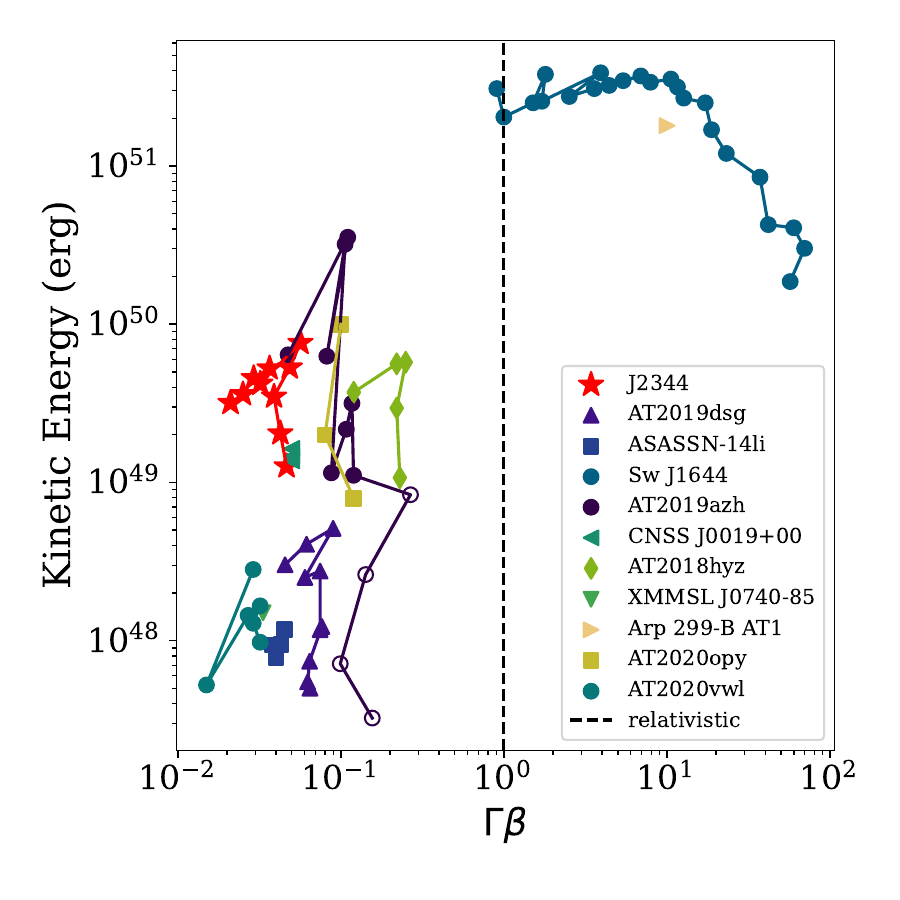}
    \caption{\textit{Left:} The variation of ambient density with distance from the black hole for TDEs with well-sampled radio lightcurves as traced by outflow modelling. \textit{Right:} The kinetic energy and velocity of the outflow produced in a selection of thermal TDEs. The equipartition corrected estimated kinetic energy is plotted for J2344 for the spherical model. In both panels J2344 is shown with red stars. J2344 appears to fit well into the population of non-relativistic TDEs in terms of energy, velocity, and ambient density (although the ambient density appears to be slightly lower than other events).  TDE data and assumed SMBH masses are from \citet{Cendes2021,Stein2021} (AT2019dsg, $M_{\mathrm{BH}}=5\times10^6$\,$M_{\mathrm{\odot}}$), \citet{Alexander2016} (ASASSN-14li, $M_{\mathrm{BH}}=1\times10^6$\,$M_{\mathrm{\odot}}$), \citet{Eftekhari2018} (Sw J1644+57, $M_{\mathrm{BH}}=1\times10^6$\,$M_{\mathrm{\odot}}$), \citet{Anderson2020} (CNSS J0019+00, $M_{\mathrm{BH}}=1\times10^7$\,$M_{\mathrm{\odot}}$), \citet{Mattila2018} (Arp 299-B AT1, $M_{\mathrm{BH}}=2\times10^7$\,$M_{\mathrm{\odot}}$), \citet{Alexander2017} (XMMSL1 J0740-85, $M_{\mathrm{BH}}=3.5\times10^6$\,$M_{\mathrm{\odot}}$), \citet{Goodwin2022} (AT2019azh, $M_{\mathrm{BH}}=3\times10^6$\,$M_{\mathrm{\odot}}$), \citet{Goodwin2023a} (AT2020opy, $M_{\mathrm{BH}}=1.12\times10^7$\,$M_{\mathrm{\odot}}$), \citet{Goodwin2023b} (AT2020vwl, $M_{\mathrm{BH}}=6.17\times10^5$\,$M_{\mathrm{\odot}}$). For J2344 we assume $M_{\mathrm{BH}}=6.3\times10^7$\,$M_{\mathrm{\odot}}$ \citep{Homan2023}.}
    
    \label{fig:tde_comparison_outflows}
\end{figure*}

The radio properties of J2344 are consistent with those of other radio-detected TDEs that launched non-relativistic outflows (Figures \ref{fig:tde_comparison} and \ref{fig:tde_comparison_outflows}). The peak radio and optical luminosities of J2344 are among the brightest of the population of events, perhaps indicating a correlation between the optical over-luminosity of this event and the luminosity of the radio emission in which more violent stream-stream collisions could result in a larger CIO. Further radio observations of optical over-luminous TDEs would enable confirmation of this trend. 

While J2344 appears to have a much flatter luminosity decay than other long-lived radio TDEs such as AT2019azh (Figure \ref{fig:tde_comparison}), this can be explained by a slightly flatter CNM density distribution in the host galaxy, as indicated by the outflow energetics described in Section \ref{sec:outflowprops}. Additionally, recent late-time radio observations of TDEs are revealing that late-time rising radio emission is relatively common in TDEs, regardless of whether early-time radio emission was detected or not \citep{Cendes2023}. Therefore a flattening or even re-brightening of the radio lightcurve of J2344 would not be unexpected for a TDE origin, where renewed brightening might be due to a second outflow launched by a different process. Both the timescales of the radio rise and decay with respect to the optical flare are broadly consistent with the TDE population. 
The predicted energy and velocity of the outflow that was launched also fit well into those predicted for other non-relativistic TDEs (Figure \ref{fig:tde_comparison_outflows}). 

We note that the TDE ASASSN-14li is a well-known TDE to have occurred in an LLAGN \citep{Alexander2016}, so the presence of AGN-like emission lines in the optical spectrum of J2344 \citep{Homan2023} does not preclude the possibility of a TDE occurring in the galaxy.

\subsection{AGN flare interpretation}
Given the narrow emission lines in the optical spectrum of the host galaxy of J2344 \citep{Homan2023}, an AGN flare is a natural interpretation for the variability that was observed in the nucleus of the galaxy. 
AGN often show variability across the electromagnetic spectrum, including the radio \citep[e.g.][]{Valtaoja1992b}. Radio variability of AGN is most often attributed to shocks in the jets that are the source of the radio emission, as opposed to a single ejection of material that then expands \citep[e.g.][]{Marscher1985,Valtaoja1992}. Broad blue-shifted absorption lines in some AGN X-ray and/or UV spectra are attributed to AGN disk winds \citep[e.g.][]{Fabian2012}, which can be the cause of some AGN variability. The fastest of these winds, at 0.1--0.3\,c contain highly ionised gas detectable only at X-ray energies \citep{Fiore2017}, however, given the lack of X-ray absorption lines in the X-ray observations J2344, we find the observed radio flare unlikely to be driven by an AGN disk-wind. 
In AGN adiabatic shock-in-jet flare models, an adiabatic shock develops near the base of the jet, and moves downstream, producing radio emission from synchrotron radiation that peaks at lower frequencies over time \citep{Valtaoja1992}. This model requires a pre-existing jet in the system. 


Below we analyse the radio variability properties of J2344 and compare to other radio flares observed from AGN. \citet{Hovatta2008} analysed the long-term radio variability of 55 AGN between 4.8 and 230\,GHz. They observed 159 individual AGN flares and extracted the characteristics of these flares. On average, at 4.8\,GHz, they found that the median duration of a flare was 2.9\,yr and the median variability index was 0.53, where they define the variability index as
\begin{equation}\label{eq:variability}
    V = \frac{(S_{\rm{max}} - \sigma_{S_{\rm{max}}}) - (S_{\rm{min}} + \sigma_{S_{\rm{min}}})}{(S_{\rm{max}} - \sigma_{S_{\rm{max}}}) + (S_{\rm{min}} + \sigma_{S_{\rm{min}}})},
\end{equation}
where $S_{\rm{max}}$ is the maximum observed flux density with error $\sigma_{S_{\rm{max}}}$ and $S_{\rm{min}}$ is the minimum flux density with error $\sigma_{S_{\rm{min}}}$. 

Using equation \ref{eq:variability} we calculate that at 5\,GHz for J2344 $V>0.478$ and a lower limit on the flare duration to be $>2.3$\,yr. This level of variability is consistent with the variability seen in AGN flares, although the true baseline flux of J2344 is not known so the true variability index may be higher.  

However, AGN that exhibit flares usually flare multiple times \citep[e.g.][]{Pyatunina2007}. Repeated flaring would be a clear sign of AGN activity rather than a TDE \citep{Auchettl2018}. Long-term optical and infra-red Gaia and WISE monitoring of J2344 dating back to 2014 show no previous flaring activity \citep{Homan2023} and the available archival radio observations of J2344 show no sign of previous radio activity. Future longer time baseline monitoring of the galaxy to search for additional flares may help constrain the AGN-flare model. 

In Section \ref{sec:jetoutflow} we analysed the modelled outflow properties of J2344 in the case that the radio flare was a signature of a jet either launched coincident with the optical flare, or shocks in a pre-existing jet of an AGN. We find it unlikely that the radio emission is explained by shocks in a pre-existing jet, as the optical flare would be unlikely to lead the radio flare in that case as was observed, as the optical and radio emission would be produced by the same mechanism \citep[e.g.][]{Valtaoja1992}. Additionally, we disfavour a jet from a TDE explanation for the radio outflow due to the extremely high efficiency of accretion that would be required to explain the energy observed in the outflow of this event. In the case of an AGN, the accretion rate may be lower and therefore the efficiency of accretion need not be so high in order to explain the energy in the jet. In Section \ref{sec:outflowprops} we found that the timescales and energy of the outflow are more easily explained by CNM-interactions from a single ejection of material around the time of the optical flare than a continuously-powered outflow, based on the small magnitude of the increasing energy compared to the expected accretion power that might power a jet. It therefore seems more likely that the transient emission observed from J2344 was due to a TDE in a LLAGN, as opposed to AGN activity in the galaxy.  


\section{Conclusions}\label{sec:conclusion}

We present 10 epochs of detailed radio spectral observations of a radio flare discovered from the nuclear transient J2344 which was temporally coincident with an optical, X-ray, UV, and infra-red flare. Our radio observations enable us to track the outflow properties, such as radius, energy, velocity, and magnetic field strength over the 2.5\,yrs spanned by our observations. We find that the radio flare is well explained by an expanding synchrotron-emitting region. Based on the energy and evolution timescales of this outflow, we infer that it is more likely produced by a single ejection of material from the central SMBH, than by a continuous ejection of energy into a jet from accretion. 

The evolution timescales, energetics, and luminosity of the radio flare are broadly consistent with the known non-relativistic radio-emitting TDE population. The radio luminosity of the radio emission from J2344 is among the brightest of the radio-detected TDEs to date, consistent with the high optical luminosity of the optical flare also associated with this event. In the TDE scenario, the outflow properties favour a spherical outflow launched by stream-stream collisions during the debris circularisation or the unbound debris stream, but could also be explained by a mildly collimated jet launched from accretion onto the SMBH. We find no evidence of relativistic motion of the outflow. 

The level of variability and timescale of variability of the radio flare is also broadly consistent with AGN flares, but no previous flaring activity has been detected in the host galaxy over the past decade. Due to the outflow energetics requiring a single ejection of material, rather than the AGN-flare model of shock-in-jet, we conclude that it is more likely that the nuclear transient event was produced by a TDE in a switched off or low-luminosity AGN, but we cannot rule out a sudden accretion episode in a pre-existing switched off or low-luminosity AGN. 

Future observations that continue to track the decay of the radio emission will enable the circumnuclear medium density of the host galaxy to be measured further from the central SMBH as well as determining whether the decay of the radio emission behaves similar to other radio-detected TDEs. Longer time baseline monitoring campaigns to search for additional flares and to track the evolution of the source are ongoing and will be insightful in determining the true nature of this interesting transient event.

\section*{Acknowledgements}
We thank the anonymous referee for feedback which helped to improve this manuscript. This work was supported by the Australian government through the Australian Research Council’s Discovery Projects funding scheme (DP200102471). A.M. acknowledges support by DLR under the grant 50 QR 2110 (XMMNuTra). M.K. and D.H. acknowledge support from DLR grant FKZ 50 OR 2307 and 
FKZ 50 OR 2003, respectively. The Australia Telescope Compact Array is part of the Australia Telescope National Facility (\url{https://ror.org/05qajvd42}) which is funded by the Australian Government for operation as a National Facility managed by CSIRO. We acknowledge the Gomeroi people as the Traditional Owners of the Observatory site. We acknowledge ESA Gaia, DPAC and the Photometric Science Alerts Team (\url{http://gsaweb.ast.cam.ac.uk/alerts}).

\section*{Data Availability}
The radio data presented in Table \ref{tab:radio_obs} will be available in machine readable format with the online publication of this work.



\bibliographystyle{mnras}
\bibliography{bibfile} 



\appendix

\section{Radio flux density measurements}\label{sec:radiomeasurements}

\begin{table}
    \centering
    \caption{ATCA flux density measurements of J2344. Both the statistical flux density error and additional error due to interstellar scintilaltion (ISS) are given.}
    \begin{tabular}{ccc}
Date (MJD) & Frequency (GHz) & Flux density$\pm$statistical error$\pm$ISS error ($\mu$Jy)\\
\hline
\hline
59309 & 5.0 & 274$\pm$11$\pm$82\\
59309 & 6.0 & 270$\pm$19$\pm$81\\
59309 & 8.5 & 274$\pm$20$\pm$27\\
59309 & 9.5 & 269$\pm$16$\pm$27\\
\hline
59371 & 2.6 & 279$\pm$23$\pm$126\\
59371 & 5.0 & 350$\pm$13$\pm$105\\
59371 & 6.0 & 373$\pm$11$\pm$112\\
59371 & 8.5 & 409$\pm$16$\pm$41\\
59371 & 9.5 & 399$\pm$11$\pm$40\\
\hline
59492 & 2.1 & 239$\pm$45$\pm$108\\
59492 & 5.0 & 409$\pm$23$\pm$123\\
59492 & 6.0 & 503$\pm$20$\pm$151\\
59492 & 8.5 & 600$\pm$45$\pm$60\\
59492 & 9.5 & 562$\pm$26$\pm$56\\
\hline
59541 & 1.6 & 327$\pm$53$\pm$147\\
59541 & 2.6 & 558$\pm$50$\pm$251\\
59541 & 5.0 & 528$\pm$25$\pm$158\\
59541 & 6.0 & 573$\pm$21$\pm$172\\
59541 & 8.5 & 717$\pm$24$\pm$72\\
59541 & 9.5 & 773$\pm$24$\pm$77\\
59541 & 17.2 & 323$\pm$35$\pm$0\\
59541 & 16.2 & 347$\pm$26$\pm$0\\
59541 & 21.7 & 381$\pm$44$\pm$0\\
59541 & 20.7 & 261$\pm$55$\pm$0\\
\hline
59600 & 1.6 & 403$\pm$30$\pm$181\\
59600 & 2.6 & 497$\pm$32$\pm$224\\
59600 & 5.0 & 782$\pm$37$\pm$235\\
59600 & 6.0 & 603$\pm$25$\pm$181\\
59600 & 8.5 & 792$\pm$31$\pm$79\\
59600 & 9.5 & 741$\pm$31$\pm$74\\
59600 & 17.2 & 315$\pm$25$\pm$0\\
59600 & 16.2 & 369$\pm$27$\pm$0\\
59600 & 21.7 & 269$\pm$60$\pm$0\\
59600 & 20.7 & 233$\pm$22$\pm$0\\
\hline
59673 & 1.6 & 452$\pm$59$\pm$204\\
59673 & 2.6 & 547$\pm$101$\pm$246\\
59673 & 5.0 & 565$\pm$26$\pm$169\\
59673 & 6.0 & 481$\pm$45$\pm$144\\
59673 & 8.5 & 526$\pm$43$\pm$53\\
59673 & 9.5 & 621$\pm$154$\pm$62\\
\hline
59818 & 2.1 & 328$\pm$64$\pm$148\\
59818 & 5.0 & 671$\pm$21$\pm$201\\
59818 & 6.0 & 643$\pm$17$\pm$193\\
59818 & 8.5 & 552$\pm$14$\pm$55\\
59818 & 9.5 & 495$\pm$14$\pm$49\\
59818 & 17.2 & 222$\pm$15$\pm$0\\
59818 & 16.2 & 177$\pm$18$\pm$0\\
59818 & 21.7 & 211$\pm$30$\pm$0\\
59818 & 20.7 & 218$\pm$23$\pm$0\\
\hline
59924 & 2.1 & 302$\pm$50$\pm$136\\
59924 & 5.0 & 649$\pm$25$\pm$195\\
59924 & 6.0 & 609$\pm$18$\pm$183\\
59924 & 8.5 & 467$\pm$16$\pm$47\\
59924 & 9.5 & 399$\pm$18$\pm$40\\
59924 & 17.2 & 251$\pm$32$\pm$0\\
59924 & 16.2 & 320$\pm$31$\pm$0\\
59924 & 21.7 & 351$\pm$80$\pm$0\\
59924 & 20.7 & 76$\pm$19$\pm$0\\
\hline

    \end{tabular}
    \label{tab:radio_obs}
\end{table}

\begin{table}
    \centering
    \caption{Table A1 continued}
    \begin{tabular}{ccc}
Date (MJD) & Frequency (GHz) & Flux density$\pm$statistical error$\pm$ISS error ($\mu$Jy)\\
\hline
\hline
60020 & 2.1 & 320$\pm$184$\pm$144\\
60020 & 5.0 & 565$\pm$67$\pm$169\\
60020 & 6.0 & 488$\pm$33$\pm$146\\
60020 & 8.5 & 393$\pm$24$\pm$39\\
60020 & 9.5 & 329$\pm$22$\pm$33\\
60020 & 16.7 & 144$\pm$24$\pm$0\\
\hline
60101 & 2.6 & 316$\pm$44$\pm$142\\
60101 & 5.0 & 611$\pm$22$\pm$183\\
60101 & 6.0 & 616$\pm$20$\pm$185\\
60101 & 8.5 & 395$\pm$19$\pm$40\\
60101 & 9.5 & 303$\pm$18$\pm$30\\
60101 & 16.7 & 172$\pm$18$\pm$0\\
\hline
    \end{tabular}
    \label{tab:radio_obs}
\end{table}

\section{Outflow models assuming different deviation from eqiupartition}\label{sec:equideviations}

\begin{table*}
    \centering
    \caption{Equipartition modelling properties of the outflow produced by J2344 based on the synchrotron spectral fits for different assumed deviation from equipartition where $\epsilon_B$ is the fraction of energy in the magnetic field and  $\epsilon_e$ is the fraction of energy carried by the electrons.}
    \begin{tabular}{cccccccccc}
&Time (d)\footnote{Measured with respect to the optical flare date, MJD 59095.3} & log $R$ (cm) & log $E$ (erg) & $\beta$ & log $B$ (G) & log $n_e$ (cm$^{-3}$) & log $M_{\rm{ej}}$ (g) \\
\hline
\hline
&$\epsilon_e=10^{-3}$ & $\epsilon_B=0.02$ & & & & & & & \\
\hline
&214 & 16.61$\pm$0.13 &       50.73$\pm$0.18 & 0.07$\pm$0.02 &       0.83$\pm$12.83      & 1.48$\pm$1.23      & 32.33$\pm$0.22 \\
&276 & 16.68$\pm$0.11 &       50.94$\pm$0.17 & 0.07$\pm$0.02 &       0.83$\pm$10.78      & 1.48$\pm$1.05      & 32.62$\pm$0.20 \\
&397 & 16.79$\pm$0.14 &       51.17$\pm$0.19 & 0.06$\pm$0.02 &       0.77$\pm$11.76      & 1.36$\pm$1.30      & 32.92$\pm$0.23 \\
&446 & 16.95$\pm$0.10 &       51.35$\pm$0.16 & 0.08$\pm$0.02 &       0.64$\pm$6.20      & 1.08$\pm$0.96      & 32.92$\pm$0.19 \\
Spherical&505 & 17.07$\pm$0.11 &       51.51$\pm$0.17 & 0.09$\pm$0.02 &       0.53$\pm$5.29      & 0.87$\pm$1.04      & 32.95$\pm$0.20 \\
&578 & 16.87$\pm$0.22 &       51.25$\pm$0.25 & 0.05$\pm$0.03 &       0.70$\pm$15.82      & 1.21$\pm$2.03      & 33.17$\pm$0.33 \\
&723 & 17.03$\pm$0.13 &       51.35$\pm$0.18 & 0.06$\pm$0.02 &       0.51$\pm$6.16      & 0.84$\pm$1.24      & 33.16$\pm$0.22 \\
&829 & 16.99$\pm$0.14 &       51.28$\pm$0.19 & 0.05$\pm$0.02 &       0.54$\pm$7.35      & 0.90$\pm$1.37      & 33.28$\pm$0.24 \\
&925 & 16.97$\pm$0.25 &       51.18$\pm$0.28 & 0.04$\pm$0.02 &       0.52$\pm$12.11      & 0.86$\pm$2.33      & 33.31$\pm$0.37 \\
&1006 & 16.92$\pm$0.16 &       51.13$\pm$0.20 & 0.03$\pm$0.01 &       0.56$\pm$8.22      & 0.93$\pm$1.48      & 33.41$\pm$0.25 \\
\hline
&214 & 16.99$\pm$0.13 &       51.29$\pm$0.18 & 0.16$\pm$0.05 &       0.58$\pm$7.13      & 0.97$\pm$1.23      & 32.22$\pm$0.22 \\
&276 & 17.06$\pm$0.11 &       51.50$\pm$0.17 & 0.15$\pm$0.04 &       0.58$\pm$5.99      & 0.96$\pm$1.05      & 32.50$\pm$0.20 \\
&397 & 17.17$\pm$0.14 &       51.73$\pm$0.19 & 0.14$\pm$0.04 &       0.52$\pm$6.53      & 0.84$\pm$1.30      & 32.80$\pm$0.23 \\
&446 & 17.33$\pm$0.10 &       51.91$\pm$0.16 & 0.17$\pm$0.04 &       0.38$\pm$3.45      & 0.57$\pm$0.96      & 32.81$\pm$0.19 \\
Conical&505 & 17.45$\pm$0.11 &       52.07$\pm$0.17 & 0.19$\pm$0.05 &       0.27$\pm$2.94      & 0.36$\pm$1.04      & 32.86$\pm$0.20 \\
&578 & 17.25$\pm$0.22 &       51.81$\pm$0.25 & 0.12$\pm$0.06 &       0.44$\pm$8.79      & 0.70$\pm$2.03      & 33.03$\pm$0.33 \\
&723 & 17.41$\pm$0.13 &       51.91$\pm$0.18 & 0.13$\pm$0.04 &       0.26$\pm$3.42      & 0.33$\pm$1.24      & 33.03$\pm$0.22 \\
&829 & 17.36$\pm$0.14 &       51.85$\pm$0.19 & 0.11$\pm$0.04 &       0.29$\pm$4.09      & 0.39$\pm$1.37      & 33.14$\pm$0.24 \\
&925 & 17.34$\pm$0.25 &       51.75$\pm$0.28 & 0.09$\pm$0.05 &       0.27$\pm$6.73      & 0.35$\pm$2.33      & 33.16$\pm$0.37 \\
&1006 & 17.30$\pm$0.16 &       51.69$\pm$0.20 & 0.08$\pm$0.03 &       0.30$\pm$4.57      & 0.42$\pm$1.48      & 33.25$\pm$0.25 \\

\hline

&250 & 16.99$\pm$0.13 &       51.29$\pm$0.18 & 0.14$\pm$0.04 &       0.58$\pm$7.13      & 0.97$\pm$1.23      & 32.34$\pm$0.22 \\
&312 & 17.06$\pm$0.11 &       51.50$\pm$0.17 & 0.13$\pm$0.03 &       0.58$\pm$5.99      & 0.96$\pm$1.05      & 32.59$\pm$0.20 \\
&433 & 17.17$\pm$0.14 &       51.73$\pm$0.19 & 0.13$\pm$0.04 &       0.52$\pm$6.53      & 0.84$\pm$1.30      & 32.87$\pm$0.23 \\
&482 & 17.33$\pm$0.10 &       51.91$\pm$0.16 & 0.16$\pm$0.04 &       0.38$\pm$3.45      & 0.57$\pm$0.96      & 32.87$\pm$0.19 \\
&541 & 17.45$\pm$0.11 &       52.07$\pm$0.17 & 0.18$\pm$0.05 &       0.27$\pm$2.94      & 0.36$\pm$1.04      & 32.90$\pm$0.20 \\
&614 & 17.25$\pm$0.22 &       51.81$\pm$0.25 & 0.11$\pm$0.05 &       0.44$\pm$8.79      & 0.70$\pm$2.03      & 33.08$\pm$0.33 \\
&759 & 17.41$\pm$0.13 &       51.91$\pm$0.18 & 0.12$\pm$0.04 &       0.26$\pm$3.42      & 0.33$\pm$1.24      & 33.07$\pm$0.22 \\
&865 & 17.36$\pm$0.14 &       51.85$\pm$0.19 & 0.10$\pm$0.03 &       0.29$\pm$4.09      & 0.39$\pm$1.37      & 33.18$\pm$0.24 \\
&961 & 17.34$\pm$0.25 &       51.75$\pm$0.28 & 0.09$\pm$0.05 &       0.27$\pm$6.73      & 0.35$\pm$2.33      & 33.19$\pm$0.37 \\
&1042 & 17.30$\pm$0.16 &       51.69$\pm$0.20 & 0.08$\pm$0.03 &       0.30$\pm$4.57      & 0.42$\pm$1.48      & 33.28$\pm$0.25 \\

\hline
\hline

&$\epsilon_e=10^{-4}$ & $\epsilon_B=0.02$ & & & & & & & \\
\hline
&214 & 16.72$\pm$0.13 &       51.91$\pm$0.18 & 0.09$\pm$0.03 &       1.27$\pm$35.09      & 0.19$\pm$1.23      & 33.32$\pm$0.22 \\
&276 & 16.79$\pm$0.11 &       52.12$\pm$0.17 & 0.09$\pm$0.02 &       1.27$\pm$29.48      & 0.18$\pm$1.05      & 33.60$\pm$0.20 \\
&397 & 16.90$\pm$0.14 &       52.35$\pm$0.19 & 0.08$\pm$0.02 &       1.21$\pm$32.17      & 0.06$\pm$1.30      & 33.90$\pm$0.23 \\
&446 & 17.06$\pm$0.10 &       52.53$\pm$0.16 & 0.10$\pm$0.02 &       1.07$\pm$16.97      & -0.21$\pm$0.96      & 33.90$\pm$0.19 \\
Spherical&505 & 17.18$\pm$0.11 &       52.69$\pm$0.17 & 0.11$\pm$0.03 &       0.97$\pm$14.47      & -0.42$\pm$1.04      & 33.94$\pm$0.20 \\
&578 & 16.98$\pm$0.22 &       52.43$\pm$0.25 & 0.07$\pm$0.03 &       1.14$\pm$43.27      & -0.08$\pm$2.03      & 34.15$\pm$0.33 \\
&723 & 17.14$\pm$0.13 &       52.53$\pm$0.18 & 0.07$\pm$0.02 &       0.95$\pm$16.86      & -0.45$\pm$1.24      & 34.13$\pm$0.22 \\
&829 & 17.09$\pm$0.14 &       52.47$\pm$0.19 & 0.06$\pm$0.02 &       0.98$\pm$20.12      & -0.39$\pm$1.37      & 34.26$\pm$0.24 \\
&925 & 17.07$\pm$0.25 &       52.37$\pm$0.28 & 0.05$\pm$0.03 &       0.96$\pm$33.14      & -0.43$\pm$2.33      & 34.29$\pm$0.37 \\
&1006 & 17.03$\pm$0.16 &       52.31$\pm$0.20 & 0.04$\pm$0.02 &       1.00$\pm$22.48      & -0.36$\pm$1.48      & 34.38$\pm$0.25 \\
\hline
&214 & 17.09$\pm$0.13 &       52.47$\pm$0.18 & 0.20$\pm$0.06 &       1.02$\pm$19.50      & -0.32$\pm$1.23      & 33.23$\pm$0.22 \\
&276 & 17.17$\pm$0.11 &       52.68$\pm$0.17 & 0.18$\pm$0.05 &       1.01$\pm$16.38      & -0.33$\pm$1.05      & 33.50$\pm$0.20 \\
&397 & 17.28$\pm$0.14 &       52.91$\pm$0.19 & 0.17$\pm$0.05 &       0.95$\pm$17.87      & -0.45$\pm$1.30      & 33.80$\pm$0.23 \\
&446 & 17.43$\pm$0.10 &       53.10$\pm$0.16 & 0.21$\pm$0.05 &       0.82$\pm$9.43      & -0.72$\pm$0.96      & 33.82$\pm$0.19 \\
Conical&505 & 17.56$\pm$0.11 &       53.25$\pm$0.17 & 0.23$\pm$0.06 &       0.71$\pm$8.04      & -0.93$\pm$1.04      & 33.87$\pm$0.20 \\
&578 & 17.36$\pm$0.22 &       53.00$\pm$0.25 & 0.14$\pm$0.07 &       0.88$\pm$24.04      & -0.59$\pm$2.03      & 34.03$\pm$0.33 \\
&723 & 17.51$\pm$0.13 &       53.09$\pm$0.18 & 0.16$\pm$0.05 &       0.69$\pm$9.36      & -0.97$\pm$1.24      & 34.02$\pm$0.22 \\
&829 & 17.47$\pm$0.14 &       53.03$\pm$0.19 & 0.13$\pm$0.04 &       0.73$\pm$11.18      & -0.90$\pm$1.37      & 34.13$\pm$0.24 \\
&925 & 17.45$\pm$0.25 &       52.93$\pm$0.28 & 0.12$\pm$0.07 &       0.70$\pm$18.41      & -0.94$\pm$2.33      & 34.15$\pm$0.37 \\
&1006 & 17.41$\pm$0.16 &       52.87$\pm$0.20 & 0.10$\pm$0.04 &       0.74$\pm$12.49      & -0.87$\pm$1.48      & 34.24$\pm$0.25 \\

\hline

    \end{tabular}
    \label{tab:outflowprops_appendix}
\end{table*}


\bsp	
\label{lastpage}
\end{document}